\documentclass[final,5p,times,twocolumn]{elsarticle}
\usepackage[english]{babel}
\usepackage{graphicx}
\usepackage{epsfig}
\usepackage{ae}
\usepackage{subfigure}
\usepackage{url}
\usepackage{algorithm2e}
\usepackage{svn}
\usepackage{amsfonts}
\usepackage{amsmath}
\usepackage{amssymb}
\usepackage{amstext}
\usepackage{amsthm}
\usepackage{xspace}
\usepackage{pstricks}
\usepackage{pst-node}
\usepackage{pst-plot}
\usepackage{multirow}
\usepackage{lscape}
\usepackage{picins}

\begin{document}

\begin{frontmatter}

\title{On the Energy Cost of Robustness and Resiliency in IP Networks}
\author[turin]{B. Addis}
\ead{bernardetta.addis@loria.fr}
\author[milan]{A. Capone \corref{cor1}}
\ead{capone@elet.polimi.it}
\author[milan]{G. Carello}
\ead{giuliana.carello@polimi.it}
\author[milan,montreal]{L.G. Gianoli}
\ead{gianoli@elet.polimi.it}
\author[montreal]{B. Sans\`o}
\ead{brunilde.sanso@polymtl.ca}
\cortext[cor1]{Corresponding author}
\address[milan]{Politecnico di Milano, Dipartimento di Elettronica, Informazione e Bioingegneria, Italy}
\address[turin]{LORIA, Universit\'e de Lorraine, CNRS, INRIA, Nancy, France}
\address[montreal]{\'Ecole Polytechnique de Montr\'eal, D\'epartement de G\'enie \'Electrique, Canada}
\date{}

\begin{abstract}
Despite the growing concern for the energy consumption of the Internet, green strategies for network and traffic management cannot undermine the quality and the functional level normally expected from carrier networks. In particular, two very important issues that may be affected by green networking techniques are resilience to node and link failures, and robustness to traffic variations.

In this paper, we study how strategies aiming at achieving different levels of resiliency and robustness impact the efficiency of energy-aware network management approaches in saving energy.  We propose novel optimization models to minimize the energy consumption of IP networks that explicitly guarantee network survivability to failures and robustness to traffic variations. Network consumption is reduced by putting in sleep mode idle line cards and nodes according to daily traffic variations that are modeled by dividing a single day into multiple time intervals. To guarantee network survivability we consider two different schemes, dedicated and shared protection, which assign a backup path  to each traffic demand and some spare capacity on the links along the path. Robustness to traffic variations is provided through an approach that allows to tune the capacity margin on active links in order to accommodate load variations of different magnitude. Furthermore, we impose some inter-period constraints necessary to guarantee network stability and preserve device lifetime. Both exact and heuristic methods are proposed. 

Experimentations carried out on realistic networks operated with flow-based routing protocols (like MPLS) allow us to quantitatively analyze the trade-off between energy cost and level of protection and robustness. Results show that with optimal strategies significant savings, up to 30\%, can be achieved even when both survivability and robustness are fully guaranteed.
\vspace{0.3cm}

\end{abstract}

\begin{keyword}
Energy-Aware, Traffic Engineering, Network resiliency, Robust Optimization, Shared protection, Dedicated protection
\end{keyword}

\end{frontmatter}

\section{Introduction}\label{sec:introduction}

Both network operators and device manufacturers agree that the energy consumption of communications networks cannot be neglected anymore~\cite{greentouch}. According to recent estimates~\cite{lambert12}, the worldwide electricity consumption of telecom operators has grown from 150 Twh/y in 2007  to 260 TWh/y in 2012, which accounts for almost 3\% of the total worldwide consumption. 

This growing consumption has stimulated the 
development of new strategies to increase the energy efficiency of communications networks,
with particular focus on IP networks~\cite{mellah09,bolla11,vetter12}.
In this context, remarkable improvements can be obtained with energy-aware strategies for network management and traffic engineering that dynamically optimize the network configuration by putting in sleep mode some of the network components (line cards and nodes) and using the remaining active part to serve the actual traffic~\cite{amaldi13,chabarek08,mahadevan09,bianzino12c,bolla11,zeadally11}.

Basically, the aim of energy-aware management is to adjust both the network topology and the available capacity to varying traffic levels in order to keep active only the resources that are essential for the actual load. It is basically a dynamic redesign of the network that takes as input the predicted or measured traffic pattern. Management approaches differ depending on the strategy to select the sleeping/active parts, on how optimizing the network routing through the active network, and on how closely traffic variations can be followed by network reconfiguration.

It is quite evident that the best energy performance can be obtained tailoring perfectly the network capacity to the traffic level. But on the other side there are several important functions that depend on the spare capacity available in the network in case actual working conditions are different from expected. Protection techniques are widely used to guarantee the network capability of being resilient to failures of links (or device interface serving a link) and nodes. In case of failure occurrence, the affected traffic is rerouted on the surviving part of the network. To verify the possibility of accommodating this traffic on alternative paths to destination it is obviously necessary to have some spare capacity left there to cope with these anomalous situations. Also traffic can be different from what expected because of the uncertainty intrinsic in traffic estimations or possible rapid deviations that cannot be followed by monitoring and measurement techniques. Networks should be designed to be robust to these variations, and also in this case the price to pay is some spare capacity in the network to be used for compensating traffic fluctuations around the nominal value.

For these reasons, there is clearly a trade off between energy consumption and the level of resilience and robustness of the network. However, how protection techniques and robustness strategies are integrated in the energy-aware network management methodologies is fundamental to determine their energy cost. Moreover, if technologies are available for rapidly reactivating sleeping elements when needed, more energy-efficient approaches can be defined by integrating procedures to recover from failures within the network energy management framework.

The fundamental question that we tackle in this paper is whether it is possible to design a network with embedded reliability, survivability and robustness and still aiming at energy reduction. We also want to investigate what is  the energy cost of protection and robustness considering different available techniques. To this aim, we introduce a novel framework for survivable and robust energy-aware network management that
builds on our recent work on multi-period energy-aware network management in IP networks \cite{addis13a}
\footnote{Preliminary results have been presented in \cite{addis12d,addis12e,addis13b}.}.
The central idea is to introduce dedicated and shared protection into the energy-aware IP network management
models and add as well the notion of network robustness to traffic variations \cite{bertsimas11}.

This paper is organized as follows. In Section \ref{sec:related} we review state-of-the-art literature on energy-aware network management and point out the novelties of our work. In Section \ref{sec:ems} we discuss the most relevant aspects of the proposed energy-aware framework, while in Sections \ref{sec:formulation}, \ref{sec:resilience} and \ref{sec:robust} we present some MILP formulations to take into account both survivability and robustness issues. The resolution methods and computational results are extensively discussed in Section \ref{sec:methods} and \ref{sec:results}, respectively. Finally, some concluding remarks are reported in Section \ref{sec:conclusions}.

\section{Related work}\label{sec:related}

The issues concerning the optimization of Internet energy consumption were first discussed in the seminal work by Gupta and Singh~\cite{gupta03}. Since then, in the last decade, several studies have been conducted to develop efficient strategies to make the Internet greener \cite{bianzino12c,bolla11,zeadally11}.

As pointed out in ~\cite{mellah09}, green networking proposals can be classified into: (i) new technologies and architectures for energy-efficient networking devices, (ii) virtualization-based strategies, (iii) methodologies for energy-aware network management and design. The modelling framework to minimize the daily energy consumption of IP networks proposed in this paper belongs to the last class.

Preliminary studies that evaluate the applicability and effectiveness of energy-aware network management have been  presented in \cite{bolla12,baldi09,restrepo09,bianzino10,bolla11b,chiaraviglio11b,chiaraviglio12a}. It is worth pointing out that  energy-aware network management based on device sleeping represents a promising strategy to reduce network consumption \cite{chiaraviglio11b}. 

Several studies for energy-aware network management without additional requirements such as survivability or robustness have recently been presented. To the best of our knowledge, multi-period optimization with inter-period constraints has been considered only in our  two previous articles, \cite{addis12a,addis13b}, where both a MILP exact formulation, and a GRASP heuristic to put to sleep network line cards and chassis were proposed. The other proposals can be naturally categorized according to the routing scheme considered.

The per-flow routing  taken into account in this paper was adopted in \cite{chiaraviglio11a,vasic10,avallone12}. Different off-line greedy heuristics to route traffic demands and switch-off idle links and nodes are presented in~\cite{chiaraviglio11a}. Other dynamic procedures to dynamically optimize routing paths and network power consumption are instead presented in  \cite{vasic10}; the proposed methods exploit a local search scheme and assume to cope with network devices whose consumption is strongly dependant on the utilization. Other work on energy-aware network management with flow-based routing include \cite{garroppo12a,garroppo12b,athanasiou11,chu11,coiro13,galanjiménez13,giroire12}.

Shortest path routing protocols such as OSPF are considered in \cite{bianzino12b,shen12,amaldi13,lee12,cianfrani12b}. In \cite{bianzino12b} the authors proposed a distributed algorithm to put to sleep network links by exploiting link state packets exchanged by OSPF to disseminate information concerning the link loads. Link state packets are used to exchange information also in \cite{shen12}, where a centralized network management platform to adjust OSPF link weights and equal cost path splitting ratios is presented. Energy-aware link weight optimization is also used by the heuristic methods for off-line network management proposed in \cite{amaldi13,lee12}. The OSPF protocol is instead modified in \cite{cianfrani12b} to force the network to forward the packets on a restricted set of shortest path trees. In this way, an higher number of links will remain idle and thus put to sleep.

Other energy-aware network management approaches that adopt different perspectives include a restricted path MILP formulation to put to sleep network links in networks operated with a hybrid OSPF+MPLS routing protocol \cite{zhang10}, methods for energy-aware traffic engineering in Carrier Grade Ethernet networks \cite{capone12a}, strategies to switch off links according exclusively to network topology features (traffic demands are ignored) \cite{cuomo12} and MILP models for energy-aware network planning \cite{idzikowski12}.

Preserving network resiliency is a key issue in particular when related to energy consumption. \cite{sanso09} shows for the first time
the impact on network reliability when only energy-efficiency is considered in the design. Thus, network survivability requirements must be taken into account when reducing network consumption \cite{monti11a,muhammad10,cavdar10,jirattigalachote11,bao12,aldraho12,he13,wu12,francois13}. Almost all these articles focus on the WDM domain -- only \cite{aldraho12,francois13} deals with IP networks -- and they provide a set of algorithm and models for the energy-aware management of lightpaths. Backup lightpaths are used to reserve resources to implement the desired protection scheme, and network devices that are unused or carry backup lightpaths only are put to sleep.

Dedicated protection is considered in \cite{muhammad10,monti11a,jirattigalachote11}. An ILP formulation to efficiently manage the lightpaths and based on a set of precomputed paths is proposed in \cite{muhammad10}. Some heuristic algorithm to tackle the same problem are presented in \cite{monti11a,jirattigalachote11}. 
Shared protection is instead adopted in \cite{cavdar10,bao12,he13}. In \cite{he13} the authors propose an algorithm for dynamic energy-aware admission control where connections are accepted if enough spare capacity is available in the network. Heuristic algorithms to optimize the routing of a fixed set of network demands and put to sleep network devices are presented in \cite{cavdar10,bao12}. 
Our work differs from the previous ones for several aspects, including (i) the multi-period structure of our optimization problem, (ii) the routing management performed, in our case, at the IP level, (iii) the use of two different maximum-utilization thresholds that account for the two cases when a link failure has occurred or not, and (iv) the development of both exact and heuristic methods. Furthermore, the proposed heuristic algorithm can be used also as online optimization tool.

Differently from the proposals already mentioned, in \cite{wu12} network survivability is not managed by explicitly defining backup paths, but imposing a budget on the minimum reliability required by each connection. The reliability of each path is computed by considering a measure of the failure probability observed on each link. The authors propose a heuristic to minimize the power consumption by optimizing the network routing and blocking the connection that cannot be served with the required reliability.

Finally, in \cite{aldraho12} the authors present some MILP formulation to save energy while providing protection to each single network links or demands, while in \cite{francois13} a heuristic framework to compute different topology configurations by considering a single link failure protection is proposed. 

To the best of our knowledge, except in our preliminary work presented in \cite{addis13b}, the explicit management of uncertain traffic demands taking into account the consumption reduction has been  previously addressed only in \cite{coudert13}, where the authors proposed a green robust approach to exploit redundancy elimination to reduce the amount of transmitted traffic and consequently reduce the network consumption. In \cite{coudert13} the uncertainty affects the redundancy degree of each demand and the addressed optimization problem is very different from ours. More generally, in literature the management of traffic variation issues is typically faced in an indirect way by applying on-line strategies that adjust the network configuration according to the observed traffic variations. We refer the reader to \cite{bertsimas11} and \cite{ben09} for general surveys on robust optimization applied in both general and network contexts.

Finally, we refer the reader to \cite{pioro04} for a general survey on multi-period network optimization and survivable network design.

\section{Energy Management, robustness and survivability}\label{sec:ems}
\label{sec:problem}

We here present the key elements of our proposed approach and discuss their interactions and roles when we manage the system so as to reduce the energy consumption and to guarantee at the same time a certain level of protection again failures and traffic variations. This preliminary high level description of the framework is intended to give an overview of the issues that motived our mathematical models that are then presented with all their details in next section. For this reason we included at the end of this section also a visual example on a small network that can help understand more easily the impact of resiliency and robustness on energy efficiency.

\subsection{Energy management}
Given a backbone IP network composed of routers (chassis) and links (line cards), we consider the problem of planning in advance (i.e. off-line) both routing and topology configurations so as to minimize the daily network consumption, while guaranteeing the normal network operation, in terms of both QoS and resiliency to failures. Power consumption is reduced by efficiently exploiting a subset of network equipment to route traffic demands and by putting to sleep the remaining idle devices. 

\begin{figure}[t]\centering
  \includegraphics[width=7.5cm]{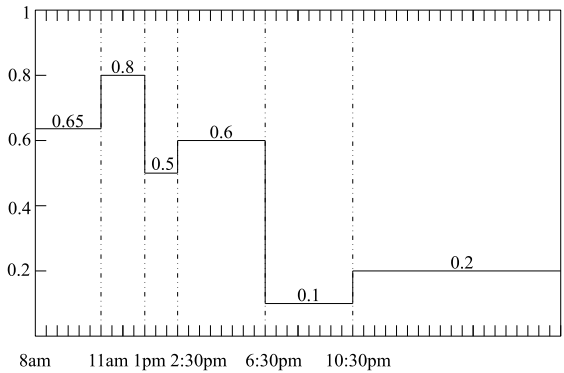}
  \caption{Traffic scenarios.}
\label{fig:scenario}
\end{figure}

To efficiently adapt the network configuration to the traffic level, we split the considered time horizon, a single day for instance, in multiple time periods, or scenarios, characterized by a given level of traffic (see Figure \ref{fig:scenario}). The time period division is performed by analyzing the daily traffic profiles (see for instance \cite{Geant}) typically observed or estimated by the network provider. The traffic demand in each time period is, for the sake of simplicity, represented by a single average estimated traffic matrix, the  \textit{traffic scenario}. The demand profile associated to the considered time horizon is cyclically repeated. 

\subsection{Demand robustness}
Due to the regular daily/weekly behaviour of the traffic profiles \cite{Geant}, network providers exploit direct \cite{netflow12} and indirect methods \cite{casas09}, to estimate traffic matrices in normal conditions. Since real traffic matrices naturally deviate around predicted values, we use robust optimization techniques to reserve enough spare bandwidth to satisfy unpredictable peaks of traffic. The basic idea is to adapt the modeling proposed  in \cite{bertsimas11}, according to which each traffic demand can vary into a close symmetric interval centered on its average traffic value, to our energy aware-problem. Next, a set of tunable parameters is used to adjust the robustness degree of the solutions by varying the total deviation allowed on each link.

In our multi-period modeling framework the consumption over the entire time horizon is minimized by jointly considering all traffic scenarios. Inter-period constraints are introduced to take into account the energy cost paid to power on a particular device and to preserve network stability. For instance, we impose a so-called card-reliability constraints to preserve the line-card lifetime by forbidding to reactivate too many times a single line card over the considered time horizon.

\subsection{Survivability}

Network resiliency to failures is provided by considering two different kind of protection schemes, i.e. dedicated and shared, which allow to protect the network in case of break down of a single link.  Given that multiple link failures and single node failures are very unlikely events, they have not been considered in this paper. Both dedicated and shared protection require the definition of a primary path and a backup path for each traffic demand. The latter is used to transmit data only after a link break down in the primary path. The two schemes have different advantages and disadvantages. Dedicated protection  reserves demand capacity on both primary and backup paths. This results in an excessive amount of reserved backup resources that, in case of single link-failure, will never be completely exploited. In shared protection, the backup paths corresponding to two link-disjoint primary paths share the same backup capacity when routed on the same link, in fact they will never be activated simultaneously. In this case the amount of backup capacity required is the maximum of the two traffic amounts. Due to the smaller amount of backup resources required, shared protection naturally allows to reduce the energy consumption. 

\subsection{A visual example}
\begin{figure*}[!p]\centering
  \includegraphics[width=16cm]{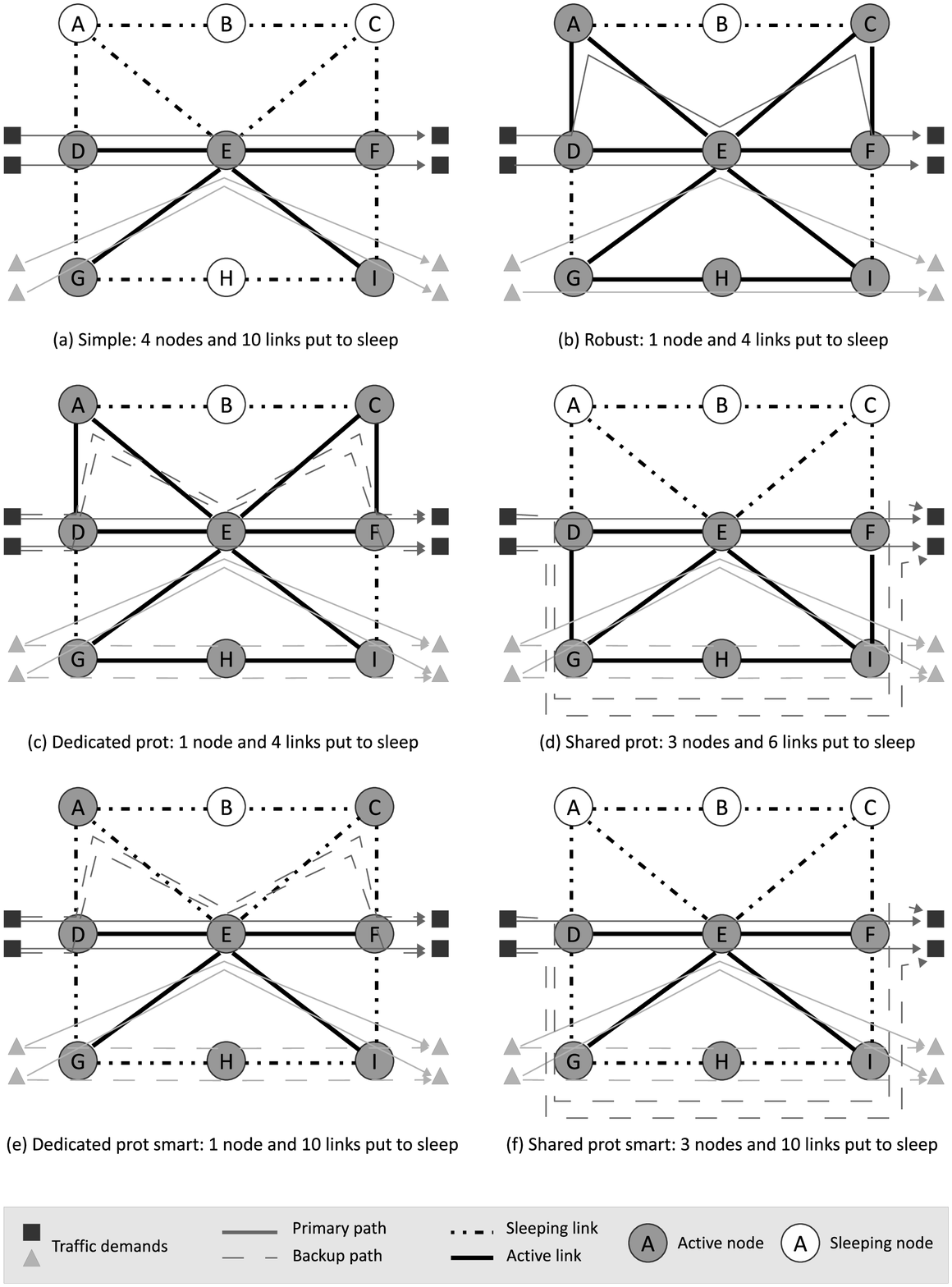}
  \caption{Energy consumption minimization vs Resilience requirements.}
\label{fig:example}
\end{figure*}

In this subsection we illustrate with an example different outcomes of our modeling framework
that depend on the protection and robustness features being introduced or not in the modeling.
They are presented in
Figures \ref{fig:example}(a), \ref{fig:example}(b), \ref{fig:example}(c) and \ref{fig:example}(d).
In the figures, link capacity is assumed to be 2 units and there are four traffic demands each requesting 1 unit of traffic.

Figure \ref{fig:example}(a) represents the simple case for which no protection schemes are implemented.
We can see that
there are 4 nodes and 10 bidirectional links to put to sleep, making this case the most energy-efficient.

When the uncertainty of traffic demands is explicitly considered  (Figure\ref{fig:example}(b)) the system cannot share any link between the four traffic demands routed in the network. Thus, 6 more links and 3 more nodes have to be activated, increasing the consumption with respect to the Simple case given above.

In case of dedicated protection (Figure \ref{fig:example}(c)), additional links and nodes have to be switched on (3 more nodes and 6 more links) to carry the backup paths. 
Note that, since each backup path has the same bandwidth requirement of the primary one, the two demands $G-I$ cannot be routed on the links already used by the backup paths of the two demands $D-F$. However, by implementing shared protection (Figure \ref{fig:example}(d)), it is possible to sensibly reduce the consumption due to protection and put to sleep 2 more nodes and 2 more links w.r.t. the dedicated case. These savings can be achieved because shared protection allows to share the backup resources on the links used by the secondary paths (links $G-H$ and $H-I$), since the two couples of demands $D-F$ and $G-I$ are satisfied two by two by link disjoint paths.

To further reduce the network power consumption, in addition to the \textit{classic} approach just described, we also investigate  a slightly modified variant, that we call \textit{smart} version, in which line cards carrying only backup paths can be put to sleep for most of the time, thus having negligible consumption. In fact, since line cards can be rapidly reactivated (in the order of milliseconds) from the sleeping state \cite{hays07}, it is reasonable to assume that those used only by  backup paths are powered only when required by the occurrence of a failure. 
It is worth pointing out that the same scheme cannot be applied to network routers because a chassis switch on requires non negligible time. Further, QoS is guaranteed by imposing a limitation on the maximum link utilization allowed. Since the occurrence of a link failure is very unlikely, we opted to include a second higher utilization threshold enable only when backup resources are exploited. Allowing the network to operate with a higher but still reasonable congestion during the very limited failure intervals we are able to further increase the energy savings.

As shown in Figures \ref{fig:example}(e) and \ref{fig:example}(f), the smart protection allows to further reduce the network consumption by switching off an additional number of links. 
When compared with the corresponding classic cases, the dedicated-smart protection 6 more links can be put to sleep, whereas with shared-smart protection the additional sleeping links are 4. Note that with the smart strategies the system is pushed to use different links to carry backup and primary paths, while with the classic ones links can be shared by primary and backup without forcing to switch on a the considered link.   
 
In the following section, we provide more details on the considered problems and describe the mathematical formulations.

\section{Reference model for energy management}\label{sec:formulation}

Let us consider a backbone IP network. Each router is composed of a chassis of capacity $\psi$ and a set of line cards each of capacity $\gamma$. Duplex links connect routers. To guarantee the  connectivity of link $(i,j)$, the same number $n_{ij}$ of line cards is available on both routers $i$ and $j$. Therefore, each link has multiple operating states based on the number of powered-on line cards. Since links have the same bandwidth in both directions, the same number of line cards has to be available on each router.
We can model the network by a symmetric directed graph $G(N,A)$, where $N$ represents the sets of chassis, and $A$ represents the bidirectional links and their associated line cards.

Furthermore, let $\pi_{ij}$ and $\bar{\pi}$ be positive real parameters representing the hourly power consumption, respectively, of a single card installed on link $(i,j)$, and of a chassis of $i$. Since the reactivation of a chassis typically causes a consumption spike, parameter $\delta$ is used to quantify the additional power consumption (normalized with respect to hourly one) associated to a chassis switching-on.

The maximum utilization level allowed on each link to guarantee the required QoS is denoted by the positive real parameter $\mu_a \in \left[0,1 \right]$. Therefore the available bandwidth on one card is given by $\mu_a \gamma$. When all the line cards connected to a given router are in the stand-by mode, the router chassis can be put to sleep too.  

Due to the multi-period nature of the addressed problem, the considered daily time horizon is split among a set $S$ of time periods $\sigma$ of duration $h_{\sigma}$. The network traffic is represented by a set of traffic demands $D$, where each demand $d\, \in \, D$ is described by a source node $o_d$, a destination node $t_d$, and the amount of demand $q_d^{\sigma}$ that has to be satisfied during period $\sigma$. Such amount is a fraction of the nominal value of demand $\rho_d$. 
Table~\ref{tab::parameters} resumes the parameters list.

We first introduce the reference MILP formulation for the multi-period energy-aware network management previously presented in \cite{addis13a}, then, in the following sections, we describe the modeling of the resiliency and robustness features. The variables used in the models are resumed in Table~\ref{tab::variables}.

\subsection{Routing constraints}
Demand $d$  routing in scenario $\sigma$ is described through binary variables $x_{ij}^{d\sigma}$, which are equal to $1$ if the routing path of demand $d$ is routed on link $(i,j)$ in scenario $\sigma$. The routing constraints (\ref{mod:balance}) represent flow conservation constraints and describe the single path unsplittable routing typically used in MPLS networks. The right hand side parameter $b_i^d$ is 1 if $i = o_d$, -1 if $i = t_d$  and 0 in all the other cases.

\begin{eqnarray}
\label{mod:balance}\sum_{(i,j) \in A} x_{ij}^{d \sigma} - \sum_{(j,i)\in A} x_{ji}^{d \sigma}= b_i^d, &
\forall \sigma \in S, i \in N, d \in D 
\end{eqnarray}

\subsection{Chassis status constraints}
The status of the chassis in each scenario is described by a binary variable $y_i^{\sigma}$, which is equal to 1 if the chassis  $i$ is on during scenario $\sigma$, and 0 otherwise. The proper value of such variables is forced by constraints~(\ref{mod:cap_chass_TON}): if a demand is routed through chassis $i$ the left hand side of equation (\ref{mod:cap_chass_TON}) is strictly positive and therefore $y_i^{\sigma}$ must be equal to one. On the other hand, if $y_i^{\sigma}$ is equal to zero, no demand can be routed through chassis $i$. The constraints guarantee also that the chassis capacity $\psi$ is not exceeded.

\begin{eqnarray}
\label{mod:cap_chass_TON}\sum_{(i,j) \in A}\sum_{d \in D} q_d^{\sigma}
x_{ij}^{d \sigma}  + & \nonumber \\
\sum_{(j,i) \in A}\sum_{d \in D}q_d^{\sigma}
x_{ji}^{d \sigma} \leq \psi y_{j}^{\sigma}, & \forall \sigma \in S, j \in N 
\end{eqnarray}

\subsection{Card and link status and capacity}
The number of active line cards on link $(i,j)$ during period $\sigma$ is represented by an integer variable $w_{ij}^{\sigma} \in [0,n_{ij}]$. As for the chassis, the status of cards is set through a set of constraints, which also guarantee  that the bandwidth available on each link, which depends on the number of active cards  $w_{ij}^{\sigma}$, is not exceeded~(\ref{mod:link_cap}). To insure a suitable level of QoS, the total used capacity is limited by a fraction $\mu_a$. Besides, constraints  (\ref{mod:simmetry}) force to keep activated the same number of line cards in both the direction of each link, so as to guarantee network stability. 

\begin{eqnarray}
\label{mod:link_cap} \sum_{d \in D} q_d^{\sigma} 
x_{ij}^{d \sigma}\, \leq\, \mu_a \gamma w_{ij}^{\sigma}, & \forall \sigma \in S, (i,j) \in A \\
\label{mod:simmetry}w_{ij}^\sigma = w_{ji}^\sigma, & \forall \sigma \in S, (i,j) \in A
\end{eqnarray}

\subsection{Chassis activation consumption constraints}
The power consumption for the reactivation of chassis $j$ at the beginning of scenario $\sigma$ is represented by continuous non~negative variable $z_j^{\sigma}$, whose value is set by means of constraint (\ref{mod_sw}).
\begin{eqnarray}
\label{mod_sw} z_{j}^{\sigma} \geq \delta \bar{\pi} \left(y_{j}^{\sigma} - y_{j}^{\sigma-1} \right), &  \forall \sigma \in S, j \in N 
\end{eqnarray}

\subsection{Card activation constraints}
To preserve the lifetime and the reliability of network equipment, a single line card cannot be switched on more than $\varepsilon$ times along an entire day. The number of activation of a card is described by auxiliary binary variable $u_{ijk}^{\sigma}$, which is equal to one if cards $k$-th linking nodes $i$ and $j$ is powered on in scenario $\sigma$. The number of card activations is limited by constraints (\ref{mod:max_switch}), while the proper value of variables $u_{ijk}^{\sigma}$ is set by constraints (\ref{mod:card_state}).

\begin{eqnarray}
\label{mod:max_switch}\sum_{\sigma \in S} u_{ijk}^\sigma \leq \varepsilon, & \forall (i,j) \in A, k \in [1,n_{ij}] \\
\label{mod:card_state}\sum_{k=1}^{n_{ij}} u_{ijk}^\sigma \geq w_{ij}^{\sigma} - w_{ij}^{\sigma-1}, & \forall \sigma \in S, (i,j) \in A 
\end{eqnarray}

\subsection{Objective function}
The objective function (\ref{mod_fob}) aims at minimizing the daily energy consumption. It is the sum of three terms taking into account, respectively, the energy consumed by the router chassis in each scenario $\displaystyle \sum_{j \in N} \bar{\pi} y_{j}^{\sigma} $, the energy used by the line cards $\displaystyle\sum_{(i,j) \in A} \pi_{ij}w_{ij}^{\sigma}$, and energy one consumed when chassis are reactivated $\displaystyle\sum_{j \in N} z_{j}^{\sigma}$.

\begin{eqnarray}\label{mod_fob}
\min\sum_{\sigma \in S}\left[h_{\sigma}\left(\sum_{j \in N} \bar{\pi} y_{j}^{\sigma}
+\sum_{(i,j) \in A} \pi_{ij}w_{ij}^{\sigma}\right)
+ \sum_{j \in N} z_{j}^{\sigma}\right] \end{eqnarray}

\begin{table}[htc]
\begin{tabular}{c|l}
\hline
$\psi$ & Chassis maximum capacity\\
$\bar{\pi}$ & Chassis power consumption\\
$\delta$ & Chassis switch on energy consumption\\
\hline
$n_{ij}$ & Number of available cards on link $(i,j)$ \\
$\gamma$ & Per card capacity\\
$\pi_{ij}$ & Card power consumption\\
\hline
$\eta$ & Maximum number of allowed card switch-on\\
$\mu_a$ & Max  primary paths arc capacity fraction\\
$\mu_b$ & Max primary and backup paths arc capacity fraction\\
\hline
$o_d$ & Origin of demand $d$\\
$t_d$ & Destination of demand $d$\\
$q_d^{\sigma}$ & Demand value on scenario $\sigma$\\
$h_{\sigma}$ & Duration of scenario $\sigma$\\
\hline
\end{tabular}
\caption{Parameters}\label{tab::parameters}
\end{table}

\begin{table}[htc]
\begin{tabular}{c|l}
\hline
$x_{ij}^d$ & Primary path routing\\
$xi_{ij}^d$ & Backup path routing\\
$y_j^{\sigma}$ & Chassis status\\
$w_{ij}^{\sigma}$ & Link/Card status\\
$u_{ijk}^{\sigma}$ & Card change of state\\
$z_j^\sigma$ & Chassis switch on energy consumption\\
$g_{ijkl}^{d \sigma}$ & Joint primary and backup path routing\\
\hline
\end{tabular}
\caption{Variables} \label{tab::variables}
\end{table}

\section{Modeling resilience}\label{sec:resilience}

Modeling a protection scheme requires the addition of both routing variables and flow conservation constraints. 

\subsection{Backup path routing constraints}
The backup paths are represented by binary variables $\xi_{ij}^{d\sigma}$, which are equal to one if backup path of demand $d$ is routed on link $(i,j)$ in scenario $\sigma$. As the variables describing the primary path $x_{ij}^{d \sigma}$, the backup path variables must satisfy the flow conservation constraints~(\ref{mod:balance_backup}).

\begin{eqnarray}\label{mod:balance_backup}
\sum_{(i,j) \in A} \xi_{ij}^{d \sigma} - \sum_{(j,i)\in A} \xi_{ji}^{d \sigma}= b_i^d &   \forall \sigma \in S,i \in N, d \in D 
\end{eqnarray}

\subsection{Link disjoint constraints}

In addition, the primary and backup path of a given demand $d$ must be link disjoint, as guaranteed by constraints~(\ref{mod:disjoint_1}) and (\ref{mod:disjoint_2}).

\begin{eqnarray}
\label{mod:disjoint_1}x_{ij}^{d \sigma} + \xi_{ij}^{d \sigma} \leq 1, & \forall \sigma \in S, (i,j) \in A, d \in D \\
\label{mod:disjoint_2}
x_{ij}^{d \sigma} + \xi_{ji}^{d \sigma} \leq 1, & \forall \sigma \in S, (i,j) \in A, d \in D 
\end{eqnarray}

\subsection{Chassis status constraints}
Chassis capacity constraints~(\ref{mod:cap_chass}) take into account the resources reserved for primary and backup paths:

\begin{eqnarray}\label{mod:cap_chass}
\sum_{(i,j) \in A}\sum_{d \in D} q_d^{\sigma}
 \left(x_{ij}^{d \sigma} + \xi_{ij}^{d \sigma}\right)  + & \nonumber \\
\sum_{(j,i) \in A}\sum_{d \in D} q_d^{\sigma} 
 \left(x_{ji}^{d \sigma} + \xi_{ji}^{d \sigma}\right) \leq \psi y_{j}^{\sigma}, & \forall \sigma \in S, j \in N
\end{eqnarray}

\subsection{Card capacity constraints}
Concerning the line card capacity, two values of the maximum card capacity fraction are considered to provide the network QoS both in case of normal network operation and of single link failure.
 If no failure occurs, the value of maximum card capacity fraction is $\mu_a$, and the constraints are inequalities~(\ref{mod:link_cap}). For the single link failure, the value of maximum card capacity fraction is $\mu_b$, which represents the maximum utilization allowed in case of failure when both primary and backup paths are used. The value $\mu_b$ is greater or equal than $\mu_a$, and it is used by the network operator to find the desired trade off between resilience and consumption. It allows the network congestion to be slightly deteriorating during the very short and unlikely periods in which a single link failure occurs, in order to achieve higher savings in normal conditions.

The capacity needed on each link is different according to the different adopted protection schemes (\ref{mod:link_cap_2}) and (\ref{mod:shared_link_cap_2}). 

\subsubsection{Dedicated protection case}
 
According to the dedicated protection scheme, the capacity constraint~(\ref{mod:link_cap_2}) states that the sum of the demands whose primary and backup paths are routed on a link cannot exceed the link available capacity.

\begin{eqnarray}
\label{mod:link_cap_2}\sum_{d \in D} q_d^{\sigma} 
\left(x_{ij}^{d \sigma} + \xi_{ij}^{d \sigma}\right)\,\leq\, \mu_b \gamma w_{ij}^{\sigma}, \forall \sigma \in S, (i,j) \in A 
\end{eqnarray}

\subsubsection{Shared protection case}
According to the shared protection scheme, the capacity on a link must be greater or equal than the sum of the demands whose primary path is routed on the considered link plus the worst backup capacity due to different failures. Such case is computed by evaluating the impact of each failure and selecting the highest one. The impact of a failure is given by the sum of the demands whose backup paths are routed on the considered link and whose primary paths fall if the considered failure occurs.
To correctly model the backup capacity to be reserved on each link, a set of binary variables $g_{ijkl}^{d \sigma}$ is introduced. A binary variable $g_{ijkl}^{d \sigma}$ is defined for each pair of links $(i,j)$ and $(k,l)$, each demand $d$ and each scenario $\sigma$, and it is equal to 1 if the demand must be rerouted on link $(i,j)$ if link $(k,l)$ fails, i.e. the traffic demand $d$ is served by a primary and a backup paths routed, respectively, on link $(i,j)$ and link $(k,l)$ in scenario $\sigma$. Constraints~(\ref{mod:bottle}) force the correct value of $g_{ijkl}^{d \sigma}$.

\begin{equation}\label{mod:bottle}
g_{ijkl}^{d \sigma} \geq x_{ij}^{d \sigma}+ \xi_{kl}^{d \sigma} - 1, \forall \sigma \in S, (i,j), (k,l) \in A,  d \in D, 
\end{equation}

Taking into account variables $g_{ijkl}^{d \sigma}$, the impact of failure  $(k,l)$ on link $(i,j)$ can be computed by constraints~(\ref{mod:shared_link_cap_2}), which allow to protect the network (i.e. to reserve enough capacity) by reserving on each link enough backup bandwidth to cope with the worst-case single link failure.

\begin{equation}\label{mod:shared_link_cap_2}
\sum_{d \in D} q_d^{\sigma} 
\left(x_{ij}^{d \sigma} + g_{klij}^{d \sigma}\right)\,\leq\, \mu_b \gamma w_{ij}^{\sigma}, \forall \sigma \in S, (i,j),(k,l) \in A
\end{equation}

\subsection{The smart consumption variant}\label{sec:smart}
The smart protection variant exploit the possibility of reactivating sleeping line cards in a few milliseconds and therefore the possibility of putting to sleep line cards that carry only backup paths during normal network operation. Such new feature of  dynamic network can be modelled by replacing constraints~(\ref{mod:link_cap_2}) with constraints~(\ref{mod2:link_cap_2}) for the dedicated protection case.

\begin{equation}
\label{mod2:link_cap_2}\sum_{d \in D} q_d^{\sigma}\left(x_{ij}^{d \sigma} + \xi_{ij}^{d \sigma}\right)\,\leq\, \mu_b \gamma n_{ij}  y_{j}^{\sigma},\;\; \forall \sigma \in S,(i,j) \in A.
\end{equation} 
Similarly, constraints~(\ref{mod:shared_link_cap_2}) must be replaced with~(\ref{mod2:shared_link_cap_2}) for the shared protection case.

\begin{equation}
\label{mod2:shared_link_cap_2}\sum_{d \in D} q_d^{\sigma}  \left(x_{ij}^{d \sigma} + g_{klij}^{d \sigma}\right)\,\leq\, \mu_b \gamma n_{ij}y_{j}^{\sigma}, \;\, \forall \sigma \in S, (i,j),(k,l) \in A
\end{equation}

Constraints (\ref{mod2:link_cap_2}) and (\ref{mod2:shared_link_cap_2}) guarantee that the total available capacity on each link ($\gamma n_{ij}$) is not exceeded by the sum of primary and backup traffic routed on it, when all the cards are switched on. H
owever, the status of cards is forced by primary paths only, as described by (\ref{mod:link_cap}). Note that the sleeping cards carrying backup paths have to be connected to an active chassis.

\section{Modeling robustness to traffic variations}\label{sec:robust}
Uncertainty may arise in the problem, if the demand amount is described by an uncertain parameter  $q_d^{\sigma}$. To deal with such uncertainty, we apply the cardinality-constrained approach proposed in~ \cite{bertsimas11}. The approach exploits the idea that all the uncertain parameters are very unlikely to assume simultaneously their worst possible value.  The uncertain parameters are assumed to vary in the interval $ \left[\overline{q}_d^{\sigma} - \hat{q}_d^{\sigma},\overline{q}_d^{\sigma} + \hat{q}_d^{\sigma} \right]$, where $\overline{q}_d^{\sigma}$ and $\hat{q}_d^{\sigma}$ represent, respectively, the expected traffic value and the maximal variation considered during period $\sigma$. In~\cite{bertsimas11} uncertainty is dealt with in such a way to guarantee than any solution is
feasible if, for each card capacity constraint associated to link $(i,j)$ and scenario $\sigma$, at most $\Gamma_{ij}^\sigma$\footnote{For the sake of conciseness, we assume parameters $\Gamma_{ij}$ to be integer. However a more general case, in which they can also continuous, can be easily dealt with, as described in~\cite{bertsimas11}.} demands, among those routed on $(i,j)$, assume their maximum value $\overline{q}_d^{\sigma} + \hat{q}_d^{\sigma}$, while all the others assume their expected one, $\overline{q}_d^{\sigma}$. 
Parameters $\Gamma_{ij}^{\sigma} \in \left[0,|D|\right] $ can be used to tune the required robustness degree by limiting the number of traffic demands that are considered uncertain. In this way, we do not limit ourself to perform a trivial worst-case optimization. Instead, by using $\Gamma_{ij}^{\sigma}$ values smaller than $|D|$, we can ignore the most unlikely realizations where all the traffic demands routed on a link $(i,j)$ assume simultaneously the maximal deviation, achieving in this way higher level of energy savings.

Uncertain parameters have an impact on constraints~(\ref{mod:link_cap})\footnote{Although uncertain parameters are present in constraint~(\ref{mod:cap_chass_TON}), uncertainty has not an impact on such constraints, as such constraints force the status of chassis variables rather than limiting the overall used capacity.}. For each such constraints a set  $U_{ij}^{\sigma}$ is defined as the set of demands which assume their maximum possible amount. The cardinality of $U_{ij}^{\sigma}$ is at most $\Gamma^{\sigma}_{ij}$. The robust counterpart of constraints~(\ref{mod:link_cap}) is:

\begin{eqnarray}\label{mod:link_cap_robust}
\sum_{d \in D} \overline{q}_d^{\sigma} x_{ij}^{d \sigma}+ 
\max_{\substack{\lbrace U_{ij}^{\sigma} \subseteq D,\,|U_{ij}^{\sigma}| \leq \lfloor \Gamma_{ij}^{\sigma} \rfloor\rbrace }} 
\left\lbrace \sum_{d \in U_{ij}^{\sigma}}\hat{q}_d^{\sigma} x_{ij}^{d \sigma} \right\rbrace\\
\nonumber  
\leq\, \mu \gamma w_{ij}^{\sigma}, \qquad \forall \sigma \in S, (i,j) \in A
\end{eqnarray}

Let $\Theta_{ij}^{\sigma}$ represents the worst case additional traffic to be considered on link $(i,j)$ during period $\sigma$, \textit{i.e.} $\Theta_{ij}^{\sigma} = \max_{\substack{\lbrace U_{ij}^{\sigma} \subseteq D,\,|U_{ij}^{\sigma}| \leq \lfloor \Gamma_{ij}^{\sigma} \rfloor\rbrace }} 
\left\lbrace \sum_{d \in U_{ij}^{\sigma}}\hat{q}_d^{\sigma} x_{ij}^{d \sigma} \right\rbrace$. 
The value of $\Theta_{ij}^{\sigma}$ can be computed through dualization.

Given a solution represented by the routing variables $\overline{x}_{ij}^{d \sigma}$, the value of $\Theta_{ij}^{\sigma}$ can be computed  solving the  following  linear programming problem:

\begin{eqnarray}\centering
\label{mod:primal_problem_fob}\Theta_{ij}^{\sigma}= \max \sum_{d \in D}\hat{q}_d^{\sigma} \overline{x}_{ij}^{d \sigma}u_{ij}^{d\sigma} & 
\end{eqnarray}
{s.t.}   
\begin{eqnarray}\centering
\label{mod:primal_problem_1}\sum_{d \in D}u_{ij}^{d\sigma} \leq \Gamma_{ij}^{\sigma} &  \\
\label{mod:primal_problem_2} 0 \leq u_{ij}^{d\sigma} \leq \ 1,  & \forall d \in D
\end{eqnarray}

Let us denote with $\epsilon_{ij}^{\sigma}$ and $l_{ij}^{d\sigma}$ the dual variables associated to constraints~(\ref{mod:primal_problem_1}) and (\ref{mod:primal_problem_2})  respectively. The dual of problem (\ref{mod:primal_problem_fob})-(\ref{mod:primal_problem_2}) is the following:
\begin{eqnarray}
\label{mod:dual_problem_fob}\min \sum_{d \in D} l_{ij}^{d\sigma}+\Gamma_{ij}^{\sigma}\epsilon_{ij}^{\sigma}& 
\end{eqnarray}
{s.t.}
\begin{eqnarray}
\label{mod:dual_problem_2} \epsilon_{ij}^{\sigma} + l_{ij}^{d\sigma} \geq \hat{q}_d^{\sigma}\overline{x}_{ij}^{d \sigma}, & \forall d \in D  \\
l_{ij}^{d\sigma} \geq 0, \epsilon_{ij}^{\sigma} \geq 0& \forall d \in D
\end{eqnarray}

According to the duality properties, the optimal primal and dual objective functions coincide and thus  the  robust constraints~(\ref{mod:link_cap_robust})  can be replaced with the following constraints:

\begin{eqnarray}
\label{mod:final_constraints_1}\sum_{d \in D} \overline{q}_d^{\sigma}x_{ij}^{d \sigma}+  \sum_{d \in D}l_{ij}^{d\sigma}+\Gamma_{ij}^{\sigma}\epsilon_{ij}^{\sigma} \leq \mu \gamma w_{ij}^{\sigma}, \\ \nonumber \forall \sigma \in S, (i,j) \in A \\ 
\epsilon_{ij}^{\sigma} + l_{ij}^{d\sigma} \geq \hat{q}_d^{\sigma} x_{ij}^{d \sigma},\forall d \in D, \forall (i,j) \in A, \forall \sigma \in S \\
l_{ij}^{d\sigma} \geq 0, \forall d \in D, \forall (i,j) \in A, \forall \sigma \in S \\
\label{mod:final_constraints_2}\epsilon_{ij}^{\sigma} \geq 0,\forall (i,j) \in A, \forall \sigma \in S \qquad \quad \,\,\,
\end{eqnarray}

It is very important to point out that the robust approach can be naturally applied to the protected case too. 

\section{Resolution methods}\label{sec:methods}
All the MILP formulations previously presented can be treated by  state-of-the-art solvers. 
In our case we experimented with CPLEX 12.5.0.0 using the AMPL modeling language and setting a resolution time limit of 1 hour. Due to the time-limit and the complexity of the models, the final solution may be sub-optimal. 

Since, for scalability reasons, the MILP formulations can be efficiently solved for instances with less than 20 nodes, 50 links and 50 demands, to solve larger instances (up to 50 nodes and 300 demands) we developed different mathematical programming heuristic techniques exploiting variants of the original MILP formulations. 

\subsection{Single time period heuristic}
The single time period heuristic (STPH) presented in \cite{addis13a} to efficiently solve the reference problem can be easily adapted to solve both the protected and the robust cases. The basic idea is to deal with each time period separately and sequentially, by solving a reduced MILP model derived from the main one, where only variables and constraints concerning the considered interval are taken into account. It is worth pointing out that some group of constraints are used to correctly evaluate the energy consumed to reactivate a chassis, and keep track of the number of switching on  each line card along all the previous periods. Card reliability constraints are respected by keeping activated all the line cards already switched on $\varepsilon$ times in the previously optimized time intervals. Since the choice of the starting period may influence the final solution, the algorithm is repeated $|S|$ times, taking at each iteration a different starting scenario.

\subsection{Path restriction}
To further speed up the single time period heuristic, we developed a new version of the algorithm, i.e. single time period heuristic with restricted paths (STPH-RP) based on the use of a pre-computed restricted set of paths assigned to each traffic demand.

To formalize the restricted-path variants of the complete MILP formulations, let $P^{d}$ represent the set of pre-computed paths assigned to demand $d$, and let $\chi_p^{d}$ and $\lambda_p^{d}$ be the binary variables equal to $1$ when path $p \in P^{d}$ is exploited by demand $d$, respectively, as primary path and backup path. Note that in the variables we neglect the scenario index $\sigma$ because in the single time period heuristic we consider one single scenario at the time.

The following set of constraints replace the flow conservation constraints (\ref{mod:balance}) to force each demand to use a single primary path:

\begin{eqnarray}\label{mod:assign_prima}
\sum_{p \in P^{d}}\chi_p^{d} = 1, & \forall d \in D
\end{eqnarray}

\noindent Similarly, for the protected case, flow conservation constraints of backup path (\ref{mod:balance_backup}) are replaced by:

\begin{eqnarray}\label{mod:assign_backup}
\sum_{p \in P^{d}}\lambda_p^{d} = 1, & \forall d \in D
\end{eqnarray}

\noindent Then, considering that:

\begin{eqnarray}
\label{equivalence_primary}x_{ij}^{d} = \sum_{p \in P^{d}: (i,j) \subset p}\lambda_p^{d}, & \forall (i,j) \in A, d \in D \\
\label{equivalence_backup}\xi_{ij}^{d} = \sum_{p \in P^{d}: (i,j) \subset p}\chi_p^{d}, & \forall (i,j) \in A, d \in D,  
\end{eqnarray}

\noindent all the original constraints have to be modified by replacing the $x_{ij}^{d}$ and $\xi_{ij}^{d}$ variables with the corresponding path-based expression.

The pre-computed paths of each demand are generated by means of the following procedure. First, for each demand $d \in D$, an LP formulation is solved to compute the maximum flow $m_{d}$ that can be routed 
from node $o_d$ to node $t_d$ when each link has unitary capacity. Then, being $\Omega$ an integer positive parameter, the precomputed paths are obtained by performing $\Omega$ iterations of the following multi-stage algorithm: (i) a random weight is assigned to each link, (ii) for each demand $d$, $m_{d}$ shortest paths (disjoint if possible) are computed by solving a minimum cost flow LP formulation, (iii) a minimum cost spanning tree is computed with the Kruskal algorithm, (iv) a single path for each demand is extracted from the links belonging to the spanning tree. At the end of the procedure $\Omega m_{d} + \Omega$ paths are available. It is worth pointing out that the two different strategies used to generate the paths allow to  find both disjoint paths to better achieve load balancing  and implement the protection schemes, and very correlated paths (with a lot of common links) to minimize the consumption.

\subsection{Warm starting}
Due to the extreme complexity of the shared protection model, we developed a procedure to warm start CPLEX (when solving both the multi-period exact model and each single period of the heuristic) with a solution rapidly obtained by solving the dedicated protection formulation with a limited time-limit. The warm-start is implemented in CPLEX using the option {\tt send$\_$statuses 2}. In case of the exact formulation the time limit is typically set to 3 minutes (with 57 minutes left to the resolution of the main model). For the heuristic the solution obtained for the previous scenario is given as input to solver, by properly initializing the variables of the considered time periods with the values of those of the previous one. The warm-start is implemented in CPLEX using the option {\tt send$\_$statuses 2}. In the heuristic, due to some AMPL code constraints, the resolution of each single period is equally split between the dedicated protection warm start and the shared protection model. Note that each feasible solution of the dedicated protection is naturally feasible for the shared protection one.

\section{Computational Results}\label{sec:results}
We performed several computational tests to evaluate both the impact of the different proposed strategies and the performance of the resolution methods. All the experiments were carried out on machines equipped with Intel i7 processors with 4 core and multi-thread 8x, and 8Gb of RAM. Test-bed and instance characteristics are described in Section \ref{sec:testbed}, the behaviour of each strategy is exhaustively discussed in Section \ref{sec:rob_prot} and, finally, extensive results on the largest instances are analysed in Section \ref{sec:large}.

\renewcommand\arraystretch{1.1}

\begin{table}[!htc]
  \centering
  \caption{Overview of different network configurations}\label{tab_card_device}
  \begin{tabular}{c|c|c c}
    \hline 
          case & device       & capacity  & hourly cons. \\ 
          \hline
          $-$ & Chassis Juniper M10i & 16Gbps & 86.4 W\\
          \hline
    $alfa$ & FE 4 ports   & 400 Mbps  & 6.8 W \\
    $delta$ & OC-3c 1 port & 155 Mbps  & 18.6 W\\
    $eta$   & GE 1 port    & 1 Gbps    & 7.3 W\\
    \hline 
  \end{tabular}

\vspace{0.5cm}
\footnotesize
\tabcolsep 5pt
\caption{Test instances.}
\begin{center}
\begin{tabular}{ccccccc}

\hline

ID	& Net	     & $|N|$-$|N_c|$	& $|A|$	& $|D|$	& \textit{equip} & \textit{scenario}	\\
\hline
1	& polska	 &  12-6	        &  36	    & 15	& alfa	 & a \\
2	& polska	 &  12-6	        &  36	    & 15	& alfa	 & b \\
3	& polska	 &  12-6	        &  36	    & 15	& alfa	 & c \\
4	& polska	 &  12-6	        &  36	    & 15	& alfa	 & aver \\
5	& polska	 &  12-6	        &  36	    & 15	& delta	 & a \\
6	& polska	 &  12-6	        &  36	    & 15	& delta	 & b \\
7	& polska	 &  12-6	        &  36	    & 15	& delta	 & c \\
8	& polska	 &  12-6	        &  36	    & 15	& delta	 & aver \\
9	& polska	 &  12-6	        &  36	    & 15	& eta	 & a \\
10	& polska	 &  12-6	        &  36	    & 15	& eta	 & b \\
11	& polska	 &  12-6	        &  36	    & 15	& eta	 & c \\
12	& polska	 &  12-6	        &  36	    & 15	& eta    & aver \\
\hline                                                          
13	& nobel-ger  &  17-9	        &  42	    & 21	& alfa	 & a \\
14	& nobel-ger  &  17-9	        &  42	    & 21	& alfa	 & b \\
15	& nobel-ger  &  17-9	        &  42	    & 21	& alfa	 & c \\
16	& nobel-ger  &  17-9	        &  42	    & 21	& alfa	 & aver \\
17  & nobel-ger  &  17-9	        &  42	    & 21	& delta	 & a \\
18	& nobel-ger  &  17-9	        &  42	    & 21	& delta	 & b \\
19	& nobel-ger  &  17-9	        &  42	    & 21	& delta	 & c \\
20	& nobel-ger  &  17-9	        &  42	    & 21	& delta	 & aver \\
21	& nobel-ger  &  17-9	        &  42	    & 21	& eta	 & a \\
22	& nobel-ger  &  17-9	        &  42	    & 21	& eta	 & b \\
23	& nobel-ger  &  17-9	        &  42	    & 21	& eta	 & c \\
24	& nobel-ger  &  17-9	        &  42	    & 21	& eta    & aver \\
\hline                                                          
25	& nobel-eu   &  28-14	        &  82	    & 90	& alfa	 & a \\
26	& nobel-eu   &  28-14	        &  82	    & 90	& alfa	 & b \\
27	& nobel-eu   &  28-14	        &  82	    & 90	& alfa	 & c \\
28	& nobel-eu   &  28-14	        &  82	    & 90	& alfa	 & aver \\
29	& nobel-eu   &  28-14	        &  82	    & 90	& delta	 & a \\
30	& nobel-eu   &  28-14	        &  82	    & 90	& delta	 & b \\
31	& nobel-eu   &  28-14	        &  82	    & 90	& delta	 & c \\
32	& nobel-eu   &  28-14	        &  82	    & 90	& delta	 & aver \\
33	& nobel-eu   &  28-14	        &  82	    & 90	& eta	 & a \\
34	& nobel-eu   &  28-14	        &  82	    & 90	& eta	 & b \\
35	& nobel-eu   &  28-14	        &  82	    & 90	& eta	 & c \\
36	& nobel-eu   &  28-14	        &  82	    & 90	& eta    & aver \\
\hline  
37	& germany    &  50-25	        &  176	    & 182	& alfa	 & a \\
38	& germany    &  50-25	        &  176	    & 182	& alfa	 & b \\
39	& germany    &  50-25	        &  176	    & 182	& alfa	 & c \\
40	& germany    &  50-25	        &  176	    & 182	& alfa	 & aver \\
41	& germany    &  50-25	        &  176	    & 182	& delta	 & a \\
42	& germany    &  50-25	        &  176	    & 182	& delta	 & b \\
43	& germany    &  50-25	        &  176	    & 182	& delta	 & c \\
44	& germany    &  50-25	        &  176	    & 182	& delta	 & aver \\
45	& germany    &  50-25	        &  176	    & 182	& eta	 & a \\
46	& germany    &  50-25	        &  176	    & 182	& eta	 & b \\
47	& germany    &  50-25	        &  176	    & 182	& eta	 & c \\
48	& germany    &  50-25	        &  176	    & 182	& eta    & aver \\
\hline 
\end{tabular}
\end{center}
\label{tab:instances}
\end{table}

\subsection{The test-bed}\label{sec:testbed} We tested both exact and heuristic methods using four network topologies provided by the SND Library (SNDLib) \cite{orlowski10}, i.e. {\tt polska}, {\tt nobel-germany}, {\tt nobel-eu} and {\tt germany}. 

The summary of the instances features is reported in Table \ref{tab:instances}, where columns $ID$, $Net$, $|N|$-$|N_c|$, $|A|$, $|D|$, \textit{equip} and \textit{scenario} represent the instance label, the network topology, the number of nodes and core nodes, the number of unidirectional link, the number of traffic demands, the equipment configuration and the traffic scenario, respectively.

In each test instance all routers are assumed to be equipped with the same type of chassis and the same type of cards. However, we experimented with three different configuration cases, \textit{alfa}, \textit{delta}, and \textit{eta}, wherein the chassis technology is always the same, while the type of cards is varied (but the same technology is used for all the cards in a given instance). Chassis and card details are reported in Table~\ref{tab_card_device}. The network nodes are equally and randomly divided between core routers and edge routers. Notice that only core routers can be put to sleep, since they are neither source nor destination of any traffic demand. 

Traffic matrices have been derived by those provided by the SNDLib. The nominal values $\rho_d$ have been computed by scaling the SNDLib matrices with a fixed parameter $\varpi_{\mu_a}^{\mu_b}$. The chosen value of $\varpi_{\mu_a}^{\mu_b}$ is the highest value such that the matrix obtained multiplying the SNDLib values by $\varpi_{\mu_a}^{\mu_b}$ can be routed in the real full active network with protection (dedicated or shared), while respecting the maximum utilization in normal conditions $\mu_a$, and the maximum utilization in failure conditions $\mu_b$. In the majority of our tests we used matrices scaled for $\varpi_{50\%}^{85\%}$ computed by considering dedicated protection. That is, we used $\mu_a$ (link max-utilization due to primary paths) equal to $50\%$ and  $\mu_b$ (link max-utilization due to both primary and backup paths) equal to $85\%$. 

\begin{figure*}[!th]\centering
  \includegraphics[width=18cm]{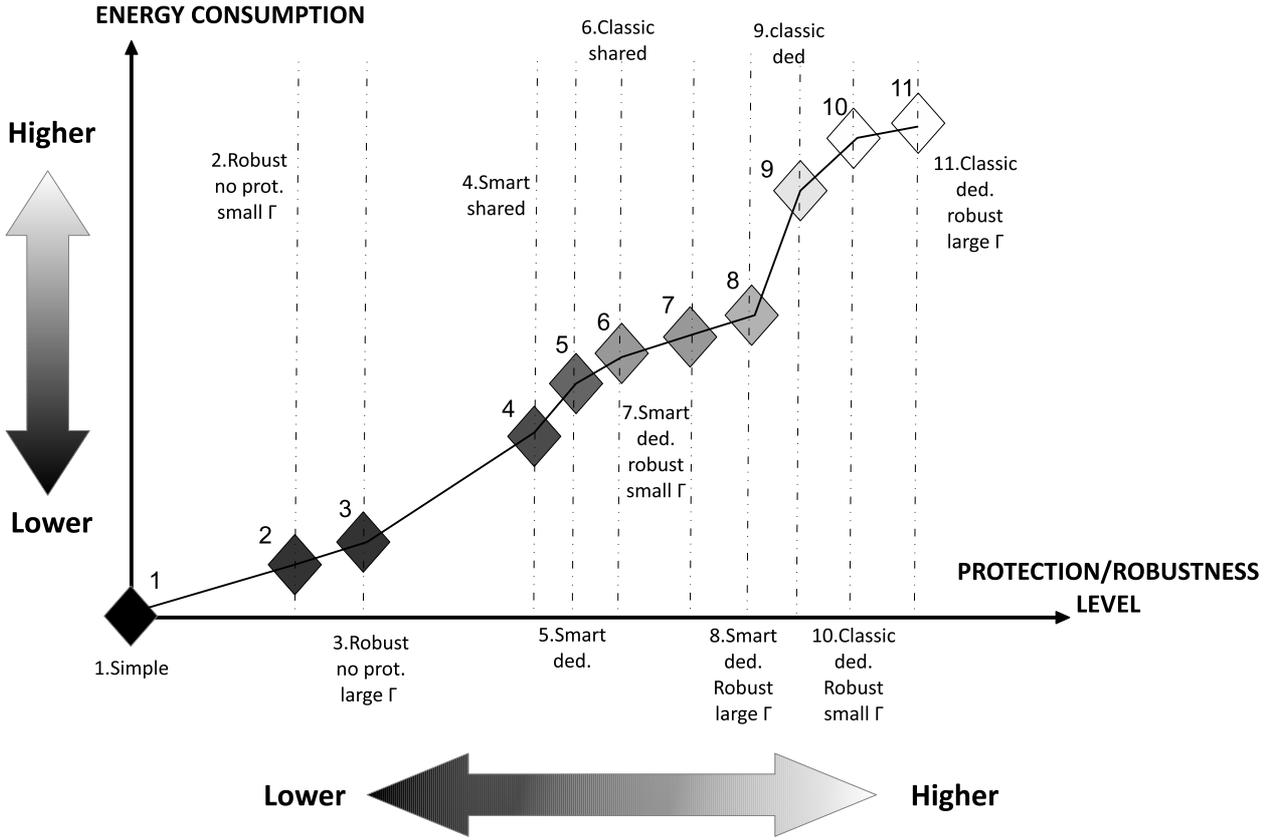}
  \caption{Savings vs Protection/Robustness.}
\label{fig:scheme}
\end{figure*}

We split a single day in six traffic periods corresponding to the following time intervals: 1) 8a.m.-11a.m., 2) 11a.m.-1p.m., 3) 1p.m.-2.30p.m., 4) 2.30p.m.-6.30p.m., 5) 6.30p.m.-10.30p.m., 6) 10.30p.m.-8a.m.
We experimented with four different traffic scenarios (column \emph{scenario}). The first three, i.e {\tt a}, {\tt b}, and {\tt c}, were generated by considering traffic values $q_d^{\sigma}$ distributed uniformly as a fraction $r_d^{\sigma}$ of nominal value $\rho_d$. In particular we considered $q_d^{\sigma}=r_d^{\sigma} \rho_d$, where parameter $r_d^{\sigma}$ is generated according to the uniform distribution $\mathcal{N}(\overline{r}_d^{\sigma}+\hat{r_d^{\sigma}},\overline{r}_d^{\sigma}-\hat{r_d^{\sigma}})$. The average values $\overline{r}_d^{\sigma}$ were chosen according to the traffic profile of Figure~\ref{fig:scenario}, the variation $\hat{r_d^{\sigma}}$ is chosen as $0.2$\footnote{Negative values are rounded up to value zero.}. 
 
In the fourth traffic scenario, namely {\tt aver} (see Table \ref{tab:instances}), all the $r_d^{\sigma}$ are equal to the average values $\overline{r}_d^{\sigma}$. This last scenario was used to compare with the robust approaches. 

To evaluate the performance of the robust approaches, we experimented with uncertainty sets of different sizes, i.e. by setting $\hat{r_d^{\sigma}}$ = 0.05,\,0.10,\,0.15,\,0.20, and by tuning the robustness degree of the solutions, i.e. by varying each $\Gamma_{ij}^{\sigma}$ from 0 (no robustness) to 5 (high level of robustness in the considered instances).

As for the remaining parameters, we set $\delta$ (chassis switching-on normalized consumption) equal to 0.25, $\varepsilon$ (switching-on limit) equal to 1, and $n_{ij}$ (number of cards in link $(i,j)$) equal to 2 for each link.

\subsection{Savings vs. Protection/Robustness}\label{sec:rob_prot}
First we aim at pointing out the impact of the different features provided to the network according to the protection/robustness strategy considered, i.e. \textit{simple}, \textit{robust}, \textit{dedicated-classic}, \textit{shared-classic} \textit{dedicated-smart}, \textit{shared-smart}, \textit{robust plus dedicated-classic} and  \textit{robust plus dedicated-smart}. The expected trade-off between energy savings and network survivability/robustness is reported in Figure \ref{fig:scheme}.
Starting from the simple energy-aware problem with no protection and no robustness, we expect the energy consumption of the network to gradually increase  if we increase the protection/robustness (P/R) level. At the first P/R level we put the \textit{robust} approach with no protection, which, by varying the robustness parameters $\Gamma_{ij}^{d\sigma}$  and the size of the uncertainty intervals $\hat{r_d^{\sigma}}$, allows to allocate additional resources to cope with traffic variations. Then, we find, in sequence,  the \textit{shared-smart} strategy and the \textit{dedicated-smart} one. Although shared protection guarantees the same degree of survivability of the dedicated one, with single link failures, we consider it less conservative because it produces solutions with, in general, less spare capacity available. Clearly, the larger  the spare capacity, the higher the capability of the network to react to failures and other unexpected events. Moving towards the right side of the graph, we first meet the \textit{shared-classic} and the \textit{dedicated-classic} strategies, and finally the two \textit{robust plus dedicated} ones. Classic schemes are considered more conservative that the smart ones because all the backup capacity is kept constantly activated. It is worth to notice that, in term of consumptions, the possibility of switching off the backup links is considered more effective than the switching from the simple dedicated protection to the complex shared one (it will be confirmed by the following results).  \textit{Robust plus shared} strategies are not reported due to the excessive computational effort required, that does not allow to efficiently solve even the smaller instances. 

To confirm the expected behaviour we considered twelve instances associated to the smallest network,  {\tt polska}, and solved the MILP formulation of each problem with a time limit of one hour.  The effectiveness of the computing methods, i.e. computing times, solution optimality, absolute savings, are evaluated, as well.

\begin{table*}[!htb]
\centering
\tabcolsep 3.7pt
\scriptsize
\begin{tabular}{rccccccccccccccccc}
& & \multicolumn{14}{c}{\textbf{\textit{Exact model - Robust approach with no protection}}} \\
\cline{4-18}
& & & \multicolumn{3}{c}{\textbf{\textit{$\Gamma = 0$}}} & \multicolumn{3}{c}{\textbf{\textit{$\Gamma = 1$}}} &  \multicolumn{3}{c}{\textbf{\textit{$\Gamma = 2$}}} & \multicolumn{3}{c}{\textbf{\textit{$\Gamma = 3$}}} & \multicolumn{3}{c}{\textbf{\textit{$\Gamma = 4$}}} \\ 
\hline
\textbf{\textit{ID}} & \textbf{\textit{$\hat{r}$}} & \textbf{\textit{TL}} & \textbf{\textit{$\%E_c$}} & $\%_{infeas}$ &	$Max_{dev}$ & \textbf{\textit{$\%E_c$}} & $\%_{infeas}$ &	$Max_{dev}$ & \textbf{\textit{$\%E_c$}} & $\%_{infeas}$ &	$Max_{dev}$ & \textbf{\textit{$\%E_c$}} & $\%_{infeas}$ &	$Max_{dev}$ & \textbf{\textit{$\%E_c$}} & $\%_{infeas}$ &	$Max_{dev}$ \\ 
\hline
4 & 0.05 & 1h & 60,6\% & 42,33\% & 6,57\% & 60,7\% & 9,91\% & 0,78\% & 60,9\% & 0,00\% & 0,00\% & 60,9\% & 0,00\% & 0,00\% & 60,9\% & 0,00\% & 0,00\%    \\ 
4 & 0.10 & 1h & 60,6\% & 81,69\% & 16,32\% & 60,9\% & 10,18\% & 2,56\% & 60,9\% & 0,09\% & 1,73\% & 60,9\% & 0,09\% & 1,73\% & 60,9\% & 0,00\% & 0,00\%  \\ 
4 & 0.15 & 1h & 60,6\% & 92,32\% & 24,64\% & 60,9\% & 15,27\% & 9,22\% & 61,4\% & 0,27\% & 4,76\% & 61,4\% & 0,27\% & 4,76\% & 61,4\% & 0,00\% & 0,00\%  \\ 
4 & 0.20 & 1h & 60,6\% & 95,60\% & 32,30\% & 60,9\% & 32,89\% & 19,19\% & 62,4\% & 0,04\% & 1,26\% & 62,4\% & 0,04\% & 1,26\% & 63,4\% & 0,00\% & 0,00\% \\ 
8 & 0.05 & 1h & 50,6\% & 40,80\% &  6,24\% & 50,8\% & 10,48\% &  0,78\% & 51,0\% & 0,00\% & 0,00\% & 51,0\% & 0,00\% & 0,00\% & 51,0\% & 0,00\% & 0,00\% \\ 
8 & 0.10 & 1h & 50,6\% & 78,39\% & 14,83\% & 51,0\% & 1,29\% &  0,61\% & 51,0\% & 0,00\% & 0,00\% & 51,0\% & 0,00\% & 0,00\% & 51,0\% & 0,00\% & 0,00\%  \\ 
8 & 0.15 & 1h & 50,6\% & 88,64\% & 24,24\% & 51,0\% & 9,49\% &  8,85\% & 51,5\% & 0,46\% & 4,89\% & 51,5\% & 0,46\% & 4,89\% & 51,6\% & 0,00\% & 0,00\%  \\ 
8 & 0.20 & 1h & 50,6\% & 93,95\% & 33,59\% & 51,0\% & 27,62\% & 19,41\% & 52,6\% & 0,05\% & 1,10\% & 52,6\% & 0,05\% & 1,10\% & 53,2\% & 0,00\% & 0,00\% \\ 
12 & 0.05 & 1h & 60,0\% & 41,37\% &  7,05\% & 60,1\% & 4,71\% &  0,70\% & 60,3\% & 0,00\% & 0,00\% & 60,3\% & 0,00\% & 0,00\% & 60,3\% & 0,00\% & 0,00\% \\ 
12 & 0.10 & 1h & 60,0\% & 78,73\% & 15,41\% & 60,3\% & 11,23\% &  2,34\% & 60,3\% & 0,00\% & 0,00\% & 60,3\% & 0,00\% & 0,00\% & 60,5\% & 0,00\% & 0,00\%\\ 
12 & 0.15 & 1h & 60,0\% & 88,49\% & 24,41\% & 60,3\% & 6,87\% &  9,17\% & 60,7\% & 0,41\% & 4,92\% & 60,7\% & 0,41\% & 4,92\% & 60,8\% & 0,00\% & 0,00\% \\ 
12 & 0.20 & 1h & 60,0\% & 94,12\% & 35,61\% & 60,3\% & 40,13\% & 18,45\% & 61,6\% & 0,01\% & 0,20\% & 61,6\% & 0,01\% & 0,20\% & 62,5\% & 0,00\% & 0,00\%\\ 
\end{tabular}
\caption{Robustness analysis  for the solutions obtained by the robust exact model with no protection  with 1h time limit on polska instances}
\label{tab:polska_model_robust}
\tabcolsep 6.3pt
\vspace{1cm}
\begin{tabular}{rcccccccccccccc}
& & \multicolumn{11}{c}{\textbf{\textit{Exact model - Robust approach with dedicated protection}}} \\
\cline{4-15}
& & & \multicolumn{3}{c}{\textbf{\textit{$\Gamma = 0$}}} & \multicolumn{3}{c}{\textbf{\textit{$\Gamma = 1$}}} & \multicolumn{3}{c}{\textbf{\textit{$\Gamma = 3$}}} & \multicolumn{3}{c}{\textbf{\textit{$\Gamma = 5$}}} \\ 
\hline
\textbf{\textit{ID}} & \textbf{\textit{$\hat{r}$}} & \textbf{\textit{TL}} & \textbf{\textit{$\%E_c$}} & $\%_{infeas}$ &	$\Delta_{smart}^{classic}$ & \textbf{\textit{$\%E_c$}} & $\%_{infeas}$ & $\Delta_{smart}^{classic}$ & \textbf{\textit{$\%E_c$}} & $\%_{infeas}$ & $\Delta_{smart}^{classic}$ & \textbf{\textit{$\%E_c$}} & $\%_{infeas}$ & $\Delta_{smart}^{classic}$ \\ 
\hline
4 & 0.05 & 1h &  70,6\% & 96,7\% & -3,1\% & 71,4\% & 17,9\% & -3,4\% & 71,5\% & 0,5\% & -3,4\% & 71,6\% & 0,0\% & -3,5\%  \\ 
4 & 0.10 & 1h &  70,6\% & 98,7\% & -3,1\% & 71,4\% & 61,7\% & -3,2\% & 71,6\% & 1,2\% & -3,3\% & 71,8\% & 0,0\% & -3,4\%  \\ 
4 & 0.15 & 1h &  70,6\% & 99,8\% & -3,1\% & 71,6\% & 63,2\% & -3,4\% & 71,9\% & 0,7\% & -3,3\% & 72,1\% & 0,0\% & -3,3\%  \\ 
4 & 0.20 & 1h &  70,6\% & 99,7\% & -3,1\% & 71,6\% & 64,1\% & -3,2\% & 72,1\% & 3,5\% & -3,1\% & 72,6\% & 0,0\% & -3,4\%  \\ 
8 & 0.05 & 1h &  60,8\% & 95,7\% & -5,5\% & 61,8\% & 31,9\% & -6,2\% & 61,8\% & 0,4\% & -5,7\% & 62,0\% & 0,0\% & -6,0\%  \\ 
8 & 0.10 & 1h &  60,8\% & 99,1\% & -5,5\% & 61,8\% & 38,6\% & -5,7\% & 62,3\% & 1,9\% & -5,4\% & 62,3\% & 0,0\% & -5,9\%  \\ 
8 & 0.15 & 1h &  60,8\% & 99,0\% & -5,5\% & 61,8\% & 63,8\% & -5,8\% & 62,6\% & 1,8\% & -5,8\% & 63,2\% & 0,0\% & -6,2\%  \\ 
8 & 0.20 & 1h &  60,8\% & 99,8\% & -5,5\% & 62,0\% & 55,4\% & -5,5\% & 62,9\% & 2,8\% & -5,3\% & 63,4\% & 0,0\% & -5,5\%  \\ 
12 & 0.05 & 1h & 70,0\% & 91,4\% & -3,2\% & 70,9\% & 21,4\% & -3,7\% & 70,9\% & 0,9\% & -3,4\% & 71,0\% & 0,0\% & -3,7\%  \\
12 & 0.10 & 1h & 70,0\% & 97,4\% & -3,2\% & 70,9\% & 32,3\% & -3,5\% & 71,0\% & 1,0\% & -3,4\% & 71,3\% & 0,0\% & -3,7\%  \\
12 & 0.15 & 1h & 70,0\% & 99,0\% & -3,2\% & 71,0\% & 56,2\% & -3,5\% & 71,4\% & 1,5\% & -3,4\% & 71,5\% & 0,0\% & -3,5\%  \\
12 & 0.20 & 1h & 70,0\% & 99,4\% & -3,2\% & 71,0\% & 71,3\% & -3,4\% & 71,6\% & 2,1\% & -3,3\% & 71,9\% & 0,0\% & -3,5\%  \\
\end{tabular}
\caption{Robustness analysis  for the solutions obtained by the robust exact model with classic dedicated protection  with 1h time limit on polska instances}
\label{tab:polska_model_robust-ded}
\end{table*}

\subsubsection{Robust strategy}
Let us first analyze the results reported in Table \ref{tab:polska_model_robust} for the \textit{robust} case. Columns $\hat{r}$ and \textit{TL}  represent the size of the demand deviation and the resolution time-limit, respectively. Then, for each instance, column $\%E_c$ represents the ratio between the energy consumption of the optimized network and the energy consumption of the full active one.
The robustness degree of the solution is evaluated on a set of randomly generated scenarios. We generated $10^,000$ random traffic scenarios where the $r^{d\sigma}$ parameters were generated with the uniform distribution $\mathcal{N}(\overline{r}_d^{\sigma}+\hat{r_d^{\sigma}},\overline{r}_d^{\sigma}-\hat{r_d^{\sigma}})$. For each generated scenario we then tested the optimized solutions by routing the random demands and verifying the violation of the capacity robust constraints. Columns $\%_{infeas}$ and $Max_{dev}$ represent the percentage of random scenarios wherein at least one capacity constraint was violated, and the largest positive difference between the observed maximum utilization and the allowed maximum one, respectively. A solution can be considered completely robust if  $\%_{infeas}\,=$ 0\%. Results clearly show that, thanks to the robust model, the optimized solutions can be completely immunized to traffic variations by using the robust parameter $\Gamma$ equal to 4 (four demands considered uncertain on each link). Most importantly, the absolute  energy consumption increase necessary to reserve additional resources is smaller, in average, than 1\%, and, in the worst case (instance 4 with $\bar{r}$ = 0.2) equal to 2.8\%. It is worth pointing out that (i) the nominal solution ($\Gamma = 0$) is largely unreliable, with $\%_{infeas}$ 95.6\% and $Max_{dev}$ = 35.6\% in the worst case (instance 12 with $\overline{r}_d^{\sigma}$ = 0.2), (ii) the increase of $\Gamma$ produces a gradual improvement of the robustness degree, as expected, (iii) the increase of the uncertainty interval force the model to reserve more resources and reduce the potential savings. The results analysis suggests that robustness has a relatively small energetic cost and allows to greatly reduce the violations of the maximum utilization constraint due to traffic variations, which is a very crucial aspect for any off-line network management approach.

\subsubsection{Protection strategies: energy efficiency}
Results concerning the protection scheme models are reported in Tables \ref{tab:polska_model_protected}-\ref{tab:polska_STPH_smart_vs_classic}.
  In Table~\ref{tab:polska_model_protected}, the energy savings achieved by \textit{simple}, \textit{dedicated-classic}, and \textit{shared-classic} models is shown.  Column $\%E_c$ represents the ratio between the energy consumption of the optimized network and the energy consumption of the full active one. Column $gap_{opt}$ represents the gap of the final solution w.r.t. to the best lower bound computed by CPLEX. Column $gap_{simple}$ represents the relative increase of energy consumption due to the survivability requirement: it is computed as $E_c^{prot}-E_c^{simple}/E_c^{simple}$, where $\%E_c^{prot}$  and $\%E_c^{prot^{simple}}$ represent the energy consumption of the optimized network w.r.t. the full active one, for the unprotected and protected case, respectively.    
  As expected, the explicit implementation of a protection scheme increases the network energy consumption, in fact the energy-aware approaches keep activated additional resources to cope with possible failures. In the case without protection, the consumption $E_c$ varies from 50.1\% to 60.6\%, in the \textit{dedicated-classic} case network consumption is between 61.4\% and 71.4\%, with absolute and relative increase, on average, of 10\% and 20\%, respectively. By considering the more sophisticated \textit{shared-classic} protection, the consumption can be reduced, w.r.t. the \textit{dedicated-classic} case, up to 5\%. However, while for the \textit{dedicated-classic}, the model computes nearly optimal solutions within the time limit of one hour ( $gap_{opt}$ usually lower than $1\%$ and never above 3.5\%), for the \textit{shared-classic} case the gap from the best lower bound is in some instances larger than 15\% (Instances 6-7), as the model is more complex and requires a high computational effort. For this reason, in some instances the reported difference between shared and dedicated protection consumption is smaller than 1\% (Instances 6-7-11). 
  
  To overcome this problem, the single time period heuristic can be applied. Heuristic results are reported in Tables~\ref{tab:polska_model_vs_heur} and \ref{tab:polska_STPH_smart_vs_classic}.
In Table~\ref{tab:polska_model_vs_heur} we analyze the gap between exact model and STPH solutions.
Columns $Heur_{gap}$ represent the difference between the energy consumption obtained by the model and that achieved by STPH, i.e.  $E_c^{heur}-E_c^{model}$.  In Table \ref{tab:polska_STPH_smart_vs_classic}  we compare the saving improvement achieved by the smart protection solution produced by STPH w.r.t. the classic one. Columns $\Delta_{smart}^{classic}$ represent the absolute difference between the energy consumption obtained with the smart and the classic models. The time limits are reported, as well: in Tables \ref{tab:polska_model_vs_heur} and \ref{tab:polska_STPH_smart_vs_classic}, differently from Tables \ref{tab:polska_model_robust} and \ref{tab:polska_model_protected}, $TL$ represent the time limit imposed to CPLEX when solving a single time period of STPH.


\begin{table}[!tbp]
\centering
\scriptsize
\tabcolsep 3pt
\begin{tabular}{rccccccccc}
& & \multicolumn{8}{c}{\textbf{\textit{Exact model}}} \\
\cline{3-10}
& & \multicolumn{2}{c}{\textbf{\textit{simple case}}} & \multicolumn{3}{c}{\textbf{\textit{dedicated prot classic}}} & \multicolumn{3}{c}{\textbf{\textit{shared prot classic}}}  \\ 
\hline
\textbf{\textit{ID}} & \textbf{\textit{TL}} & \textbf{\textit{$\%E_c$}} & \textbf{\textit{$gap_{opt}$}} & \textbf{\textit{$\%E_c$}} & \textbf{\textit{$gap_{opt}$}} & \textbf{\textit{$gap_{simple}$}} & \textbf{\textit{$\%E_c$}} & \textbf{\textit{$gap_{opt}$}} & \textbf{\textit{$gap_{simple}$}} \\ 
\hline
1 & 1h & 60,6\% & 1,3\% & 71,4\% & 1,4\% & 17,8\% & 66,9\% & 3,6\% & 10,3\%   \\
2 & 1h & 60,5\% & 0,9\% & 71,3\% & 0,9\% & 17,8\% & 66,3\% & 4,4\% & 9,6\%    \\ 
3 & 1h & 60,3\% & 0,6\% & 71,4\% & 0,7\% & 18,4\% & 70,4\% & 8,9\% & 16,7\%   \\
5 & 1h & 50,7\% & 2,4\% & 62,2\% & 2,6\% & 22,7\% & 59,3\% & 10,1\% & 17,0\%  \\
6 & 1h & 50,1\% & 0,8\% & 61,4\% & 3,3\% & 22,7\% & 60,3\% & 15,5\% & 20,5\%  \\
7 & 1h & 50,3\% & 0,4\% & 61,7\% & 2,7\% & 22,6\% & 61,7\% & 15,8\% & 22,6\%  \\
9 & 1h & 60,0\% & 1,4\% & 70,9\% & 0,9\% & 18,1\% & 66,2\% & 3,2\% & 10,3\%   \\
10 & 1h & 59,8\% & 0,7\% & 70,7\% & 0,8\% & 18,1\% & 65,7\% & 3,6\% & 9,8\%   \\
11 & 1h & 59,7\% & 0,0\% & 70,8\% & 0,5\% & 18,6\% & 70,9\% & 11,1\% & 18,8\% \\
\end{tabular}
\caption{Comparison between simple and protected solutions obtained by solving the exact model with 1h time limit with polska instances.}
\label{tab:polska_model_protected}
\end{table}

\begin{figure*}[!ht]\centering
  \includegraphics[width=18cm]{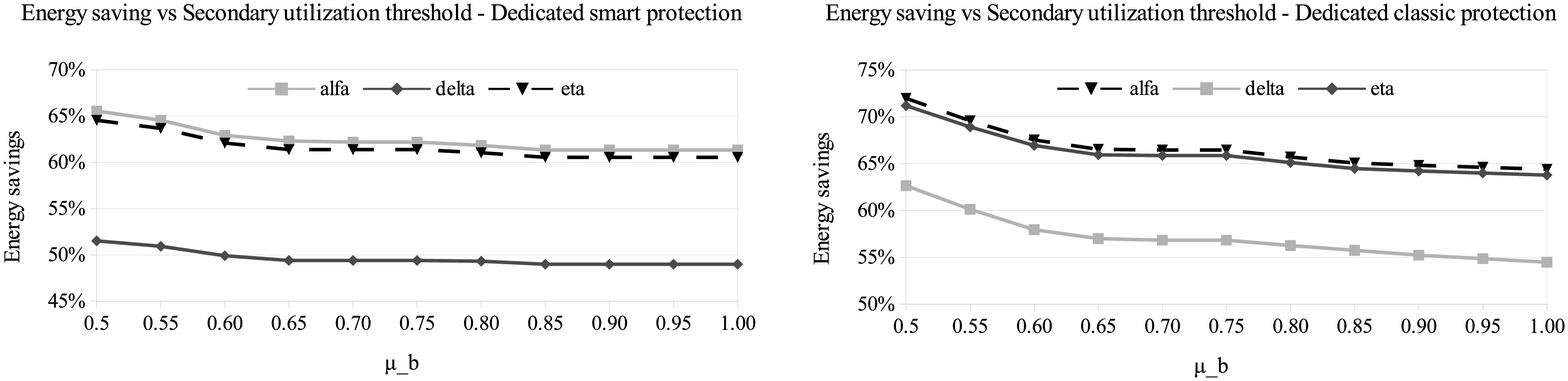}
  \caption{Analysis of the trade-off between energy savings and network congestion, obtained by adjusting the secondary utilization threshold $\mu_b$ from 0.5 to 1 when solving STPH.}
\label{fig:trade-off}
\end{figure*}

Table \ref{tab:polska_model_vs_heur} shows that, by using STPH with a time limit of 6 minutes, we reduce the energy consumption of the solutions with shared protection up to about 5\% (Instances 3-6-7-11). The difference between shared and dedicated protection for the instances for which the gap obtained solving the model is large (Instances 6-7-11) is therefore increased. Furthermore, it is worth pointing out that, even for the instances of the dedicated case solved at optimality or with a very small gap using the complete model, the gap between STPH and the formulation is very small, varying between 0.6\% and -0.2\%, negative values meaning that STPH solutions improve upon the sub-optimal solutions found by CPLEX when solving the model. Having shown the good quality of STPH algorithm solutions in the remainder of this section we report only the results obtained by solving STPH,  for practical and space reasons.

\begin{table}[!htbp]
\centering
\scriptsize
\tabcolsep 3pt
\begin{tabular}{rccccccc}
& & \multicolumn{6}{c}{\textbf{\textit{Exact model vs STPH}}} \\
\cline{3-8}
& & \multicolumn{2}{c}{\textbf{\textit{simple case}}} & \multicolumn{2}{c}{\textbf{\textit{dedicated prot classic}}} & \multicolumn{2}{c}{\textbf{\textit{shared prot classic}}}  \\ 
\hline
\textbf{\textit{ID}} & \textbf{\textit{TL$_{model}$}} & \textbf{\textit{$Heur_{gap}$}} & \textbf{\textit{$TL$}} & \textbf{\textit{$Heur_{gap}$}} & \textbf{\textit{$TL$}} & \textbf{\textit{$Heur_{gap}$}} & \textbf{\textit{$TL$}} \\ 
\hline
1 & 1h & 0,00\% & 60s & 0,1\% & 30s & -0,3\% & 360s \\
2 & 1h & 0,25\% & 60s & 0,1\% & 30s & -0,3\% & 360s \\ 
3 & 1h & 0,16\% & 60s & 0,0\% & 30s & -4,0\% & 360s \\
5 & 1h & 0,41\% & 60s & 0,6\% & 30s & -2,1\% & 360s \\
6 & 1h & 0,00\% & 60s &-0,1\% & 30s & -3,6\% & 360s \\
7 & 1h & 0,28\% & 60s &-0,2\% & 30s & -4,5\% & 360s \\
9 & 1h & 0,28\% & 60s & 0,1\% & 30s & -0,3\% & 360s \\
10& 1h & 0,28\% & 60s & 0,2\% & 30s & -0,4\% & 360s \\
11& 1h & 0,17\% & 60s & 0,1\% & 30s & -4,9\% & 360s \\
\end{tabular}
\caption{Comparison between the energy saving achieved by solving the exact model and running the single time period heuristic with different types of protection.}
\label{tab:polska_model_vs_heur}
\end{table}

The possibility of putting to sleep the line cards carrying only the backup links (\textit{smart} protection) is expected to substantially decrease the energy consumption of the network w.r.t. the \textit{classic} case. This hypothesis is clearly confirmed by the results of Table \ref{tab:polska_STPH_smart_vs_classic}, where we observe that \textit{smart} protection allows to reduce the consumption of the protected solutions (w.r.t. the total network consumption) by up to 7.1\% and 3.9\%, for the dedicated and shared case, respectively. \textit{Smart shared} produces smaller energy consumption reduction, w.r.t. the non smart case, than \textit{smart dedicated}.
The smaller reduction produced by shared protection solutions w.r.t. dedicated one is motivated by the fact that, since the first requires a minor amount of backup capacity, also lesser backup resources can be put to sleep when switching from the classic to the smart scheme. The most important result is that with the smart scheme dedicated protection can be more energy efficient than classic shared protection, 
while being less computationally expensive. 

\subsubsection{Protection strategies: congestion analysis}
Concerning the congestion, it is necessary to remind that shared protection, due to the high efficiency of the backup allocation scheme, can deal with levels of traffic that cannot be managed by the dedicated protection scheme, without violating  the maximum utilization constraints. In Table~\ref{tab_sh_nominal} the value of $\varpi_{50\%}^{85\%}$, namely the maximum values used to scale the SNDLib traffic matrix while respecting the maximum utilization constraints with $\mu_a =0.5$, $\mu_b=0.85$ and $r_d^{\sigma} = 1\, \forall d \in D, \sigma \in S$, are reported for the shared and dedicated case. Results show that shared protection allows to manage a traffic that is, on average, greater than the one managed by the dedicated case. The increase rises up to about 10\%. Therefore,  although the computational effort required by the shared case is significantly increased, shared protection scheme is worth to be implemented to reduce the network congestion.

\begin{table}[!htbp]
\centering
\scriptsize
\tabcolsep 6pt
\begin{tabular}{rcccc}
& \multicolumn{4}{c}{\textbf{\textit{STPH - Classic vs Smart }}} \\
\cline{2-5}
& \multicolumn{2}{c}{\textbf{\textit{dedicated}}} & \multicolumn{2}{c}{\textbf{\textit{shared}}}  \\ 
\hline
\textbf{\textit{ID}} & $\Delta_{smart}^{classic}$ & \textbf{\textit{$TL$}} & $\Delta_{smart}^{classic}$ & \textbf{\textit{$TL$}} \\ 
\hline
1 & -3,9\% & 30s &-1,9\%  &360s\\
2 & -3,5\% & 30s &-2,2\%  &360s\\ 
3 & -3,2\% & 30s &-1,9\%  &360s\\
5 & -7,1\% & 30s &-3,4\%  &360s\\
6 & -5,9\% & 30s &-3,9\%  &360s\\
7 & -5,5\% & 30s &-3,6\%  &360s\\
9 & -4,2\% & 30s &-2,0\%  &360s\\
10& -3,8\% & 30s &-2,3\%  &360s\\
11& -3,5\% & 30s &-2,1\%  &360s\\
\end{tabular}
\caption{Comparison between the energy saving achieved by STPH with classic and smart protection schemes.}
\label{tab:polska_STPH_smart_vs_classic}
\end{table}

To better understand the balance between network congestion and energy savings, we report in Figure~\ref{fig:trade-off} the network energy-consumption computed by varying the secondary maximum utilization threshold $\mu_b$ from 0.5 to 1. In this specific set of tests, we considered \textit{dedicated} protection and traffic matrices obtained by using $\varpi_{50\%}^{50\%}$ instead of the $\varpi_{50\%}^{85\%}$. In fact, with $\varpi_{50\%}^{85\%}$ the problem would not be feasible in case of $\mu_b < 0.85$. Figure~\ref{fig:trade-off} shows that the difference between the network consumption obtained with $\mu_b = 0.5$ and $\mu_b = 1$ varies from 4\% to 8\%. The plot clearly shows how a network provider can balance energy savings and network congestion according to his own requirements. 

\subsubsection{Joint protection and robustness}
To conclude the analysis on the {\tt polska} instances, let us analyse the results reported in Table \ref{tab:polska_model_robust-ded}, which reports about the \textit{robust dedicated} case solved with the exact formulation. As for the simple \textit{robust} case, it is possible to obtain solutions completely immunized to traffic variation ($\%_{infeas}$ = 0\%) where the power consumption is increased, on average of 1\% and in the worst case of 2.6\% (Instance 8, $\overline{r}_d^{\sigma}$ = 0.2). Furthermore, the  energy consumption reduction obtained by the smart approach w.r.t. the classic one, is about 4\% and up to 6\% (similar to the protected non robust case). 

\begin{figure*}[t]\centering
  \includegraphics[width=18cm]{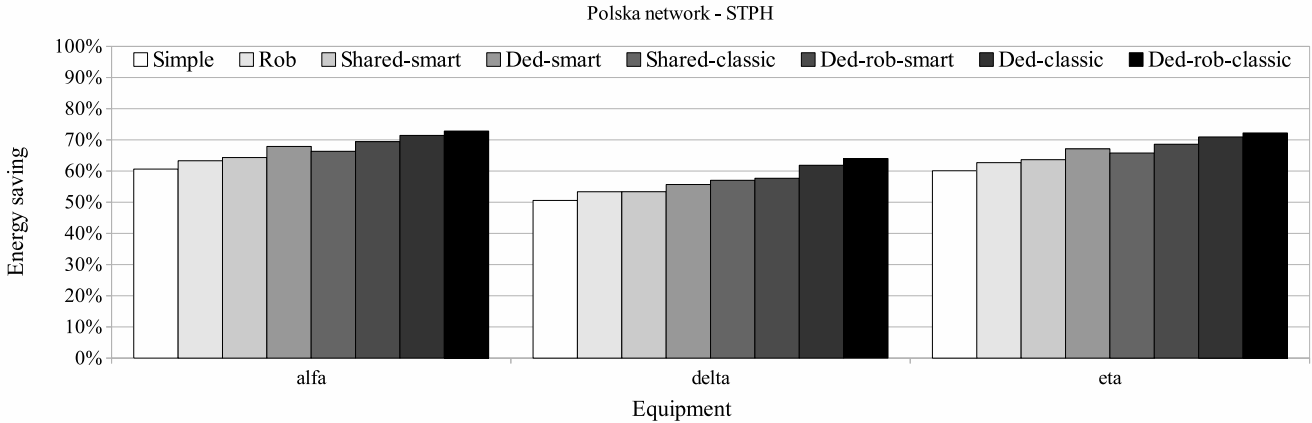}
  \caption{Energy savings achieved by STPH when implementing the different protection schemes on {\tt polska} instances.}
\label{fig:polska_general}
\end{figure*}

Finally, Figure~\ref{fig:polska_general} reports the average energy savings obtained with the three scenarios $a$, $b$, and $c$.  It is possible to observe that the final energy consumption for {\tt polska} computed by varying the protection degree follows the expected trend previously showed in Figure \ref{fig:scheme}.

\begin{table}[!ht]\footnotesize
\tabcolsep 6pt
\caption{Comparison between the efficiency of the shared protection and the dedicated protection schemes.}
\begin{center}
\begin{tabular}{crr}

\multicolumn{1}{c}{} & \multicolumn{1}{c}{$Shared$}& \multicolumn{1}{c}{$Ded$} \\
\hline
 ID   & $\varpi_{50\%}^{85\%}$	&  $\varpi_{50\%}^{85\%}$  \\
\hline                      
1-2-3-4     & 1053.7  &  941.8 \\
5-6-7-8     & 408.32  &  365.0 \\
9-10-11-12  & 2634.3  & 2354.6 \\
\hline
13-14-15-16 & 18039.3 &16451.6 \\
17-18-19-20 & 6990.5  & 6375.0 \\ 
21-22-23-24 & 45099.9 &41129.0 \\ 
\hline
\end{tabular}
\end{center}
\label{tab_sh_nominal}
\end{table}

\subsection{Largest networks}\label{sec:large}

In the second group of tests, we experimented with {\tt nobel-germany}, {\tt nobel-eu} and {\tt germany} network by running STPH or STPH-RP (the restricted path version of STPH). For comparison purposes, a restricted set of instances were tested with both procedures. STPH-RP was then used to solve instances that were too computationally  demanding to be efficiently solved in a reasonable amount of time by the simple STPH. The time limit for the single time period used to run STPH and STPH-RT are reported in Table \ref{tab:time_limit}. We report a $/$ when a given network has not been solved with the corresponding method (for instance with {\tt nobel-germany} we used only STPH). 

\begin{figure*}[t]\centering
  \includegraphics[width=18cm]{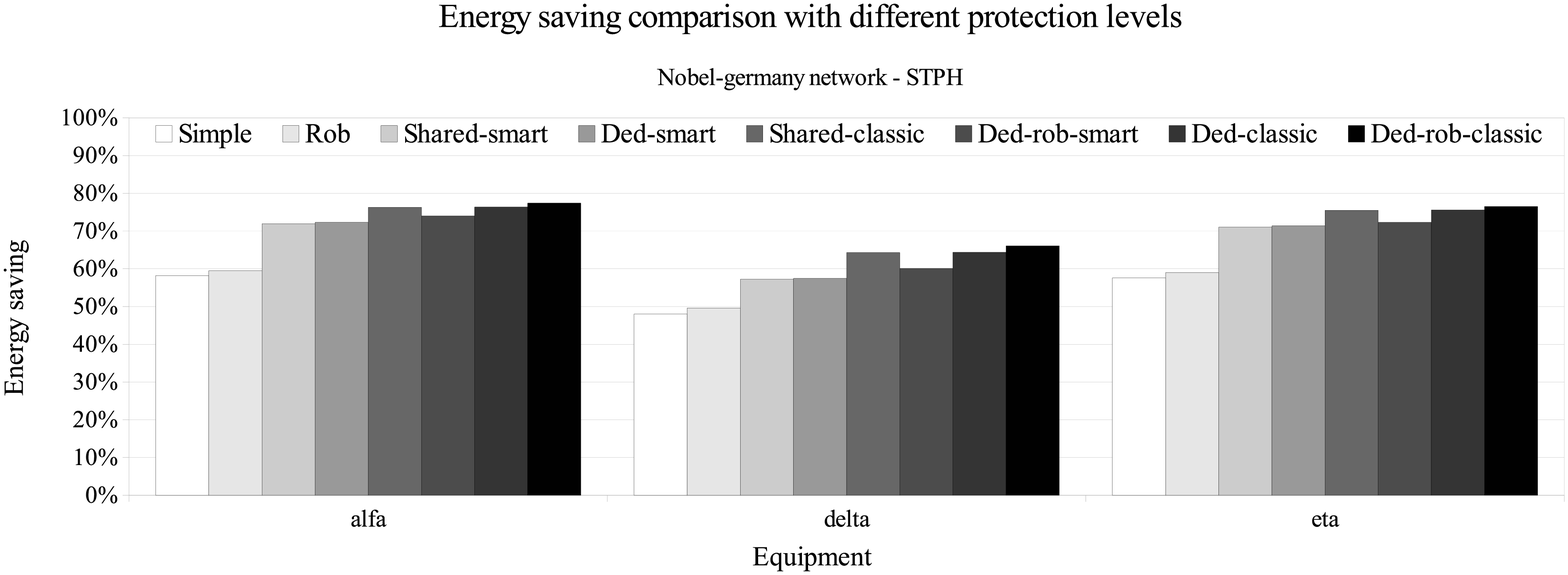}
  \caption{Energy savings achieved by STPH when implementing the different protection schemes on {\tt nobel-germany} instances.}
  \label{fig:nob_germany}
  %
  \includegraphics[width=18cm]{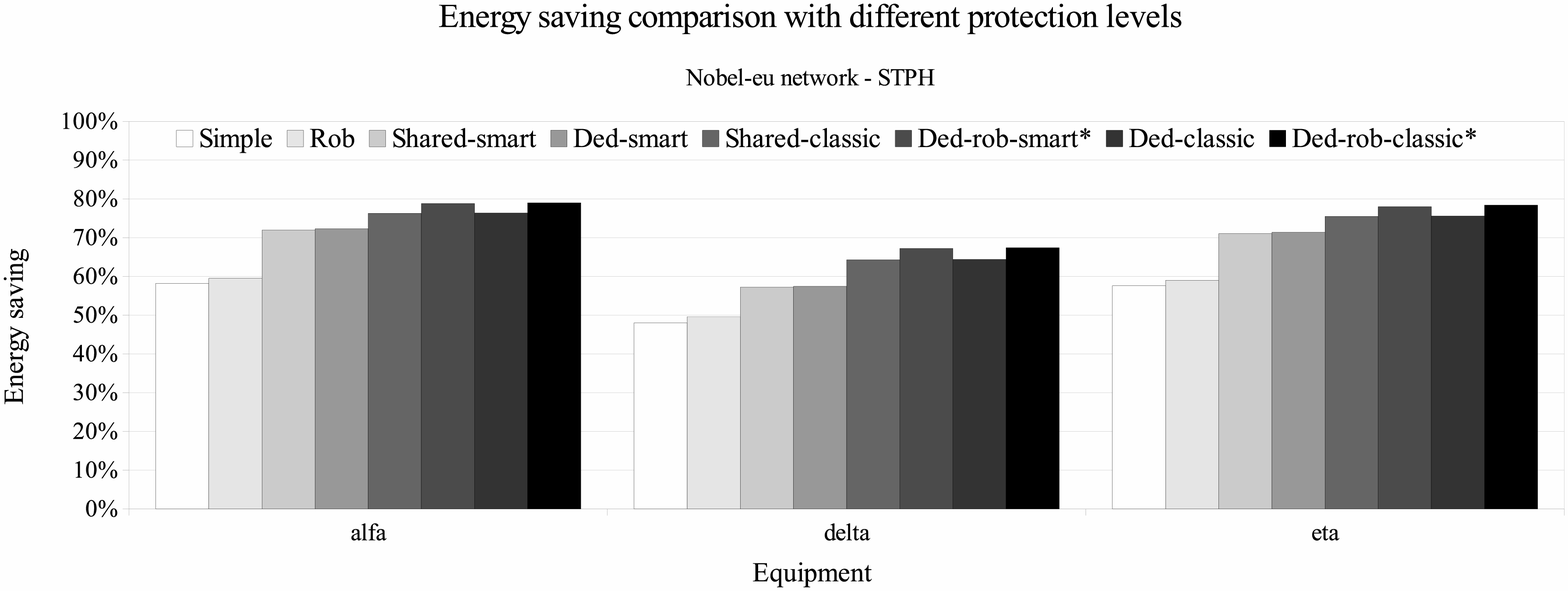}
  \caption{Energy savings achieved by STPH when implementing the different protection schemes on {\tt nobel-eu} instances. The $\ast$ in the graph legend is used for the instances solved, due to complexity issues, with STPH-RP using $\Omega=10$. }
  \label{fig:nob_eu_general}
\end{figure*}

Figure~\ref{fig:nob_germany} reports the results for the {\tt nobel-germany} network, while Figure~\ref{fig:nob_eu_general} reports the results for the {\tt nobel-eu} network.

First of all, we can observe that the consumption trend showed in Figure \ref{fig:polska_general}, is confirmed in Figures \ref{fig:nob_germany} and \ref{fig:nob_eu_general}, where we report the network consumption obtained by STPH on {\tt nobel-germany} and {\tt nobel-eu} networks considering different protection cases. The only difference that can be observed is that the energy consumption for the \textit{dedicated-smart} case is on average smaller that that one of the \textit{shared-classic} case. This can be explained as the solver is not able to efficiently solve the shared protection model, even for single period, and to obtain a small gap  when the instance dimensions increase. It is worth to note that, in some tests, the solution computed by the warm start procedure cannot be improved by the solver within the chosen time-limit. Besides, due to memory limits (8GB of RAM), the shared protection instances could not even be initialized for the  {\tt nobel-eu} and {\tt germany} networks. Thus,  as a solution feasible for the dedicated problem is naturally feasible for the shared one the solutions obtained by solving the \textit{dedicated} problem are applied also for the shared case. By solving STPH, the consumption difference between the {\tt simple} case and  the most protected one, i.e. the {\tt dedicated-classic robust}, is around 20\% for both {\tt nobel-germany}, {\tt nobel-eu} networks.

\begin{figure*}[t]\centering
  \includegraphics[width=18cm]{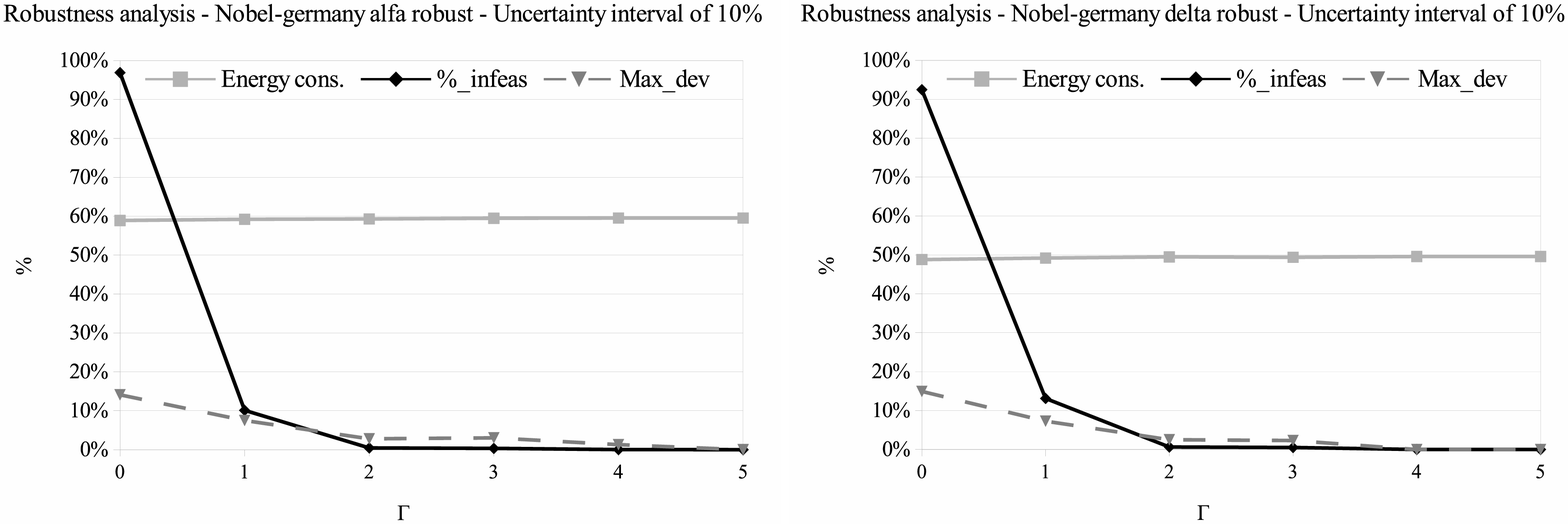}
  \label{fig:nbger_robust_analysis}
  \caption{Energy savings achieved by STPH when implementing the robust scheme on {\tt nobel-germany} instances.}
  \includegraphics[width=18cm]{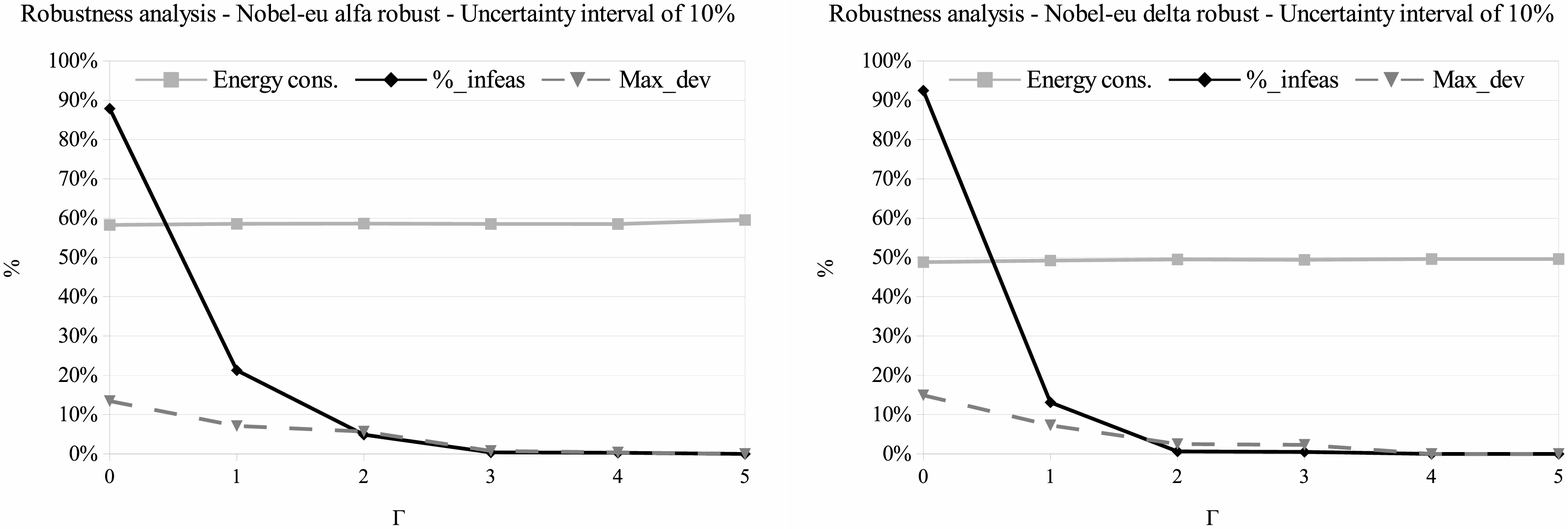}
  \label{fig:nb_eu_robust_analysis}
  \caption{Energy savings achieved by STPH when implementing the robust scheme on {\tt nobel-eu} instances.}
\end{figure*}

In Figure \ref{fig:nb_eu_robust_analysis},  the network energy consumption, the infeasibility degree and the maximum threshold overrun  are represented  for the robust case (in Table \ref{tab:polska_model_robust}, the last two values are indicated as $\%_{infeas}$ and $Max_{dev}$ ). The four graphics clearly prove that our approach allow to efficiently manage traffic variations without considerably increasing the network consumption (increase lower than 2\% for solution completely immunized). It is also worth pointing out that by simply increasing $\Gamma$ from 0 (no robustness) to 1, we are already able to substantially immunize the solution, with $\%_{infeas}$ improved from 90\% to around 15\%. The results for the $eta$ case are very similar and therefore not reported.

\begin{figure}[t]\centering
  \includegraphics[width=9cm]{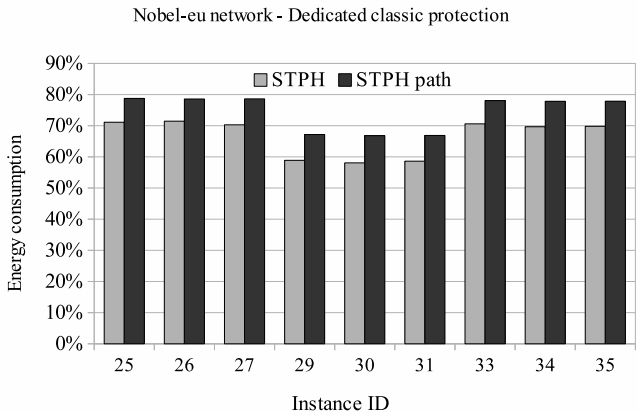}
  \caption{Energy saving comparison between STPH and STPH-RP on {\tt nobel-eu} network with \textit{dedicated classic} protection.}
\label{fig:STPH_vs_STPH-RP}
\end{figure}

For the \textit{dedicated robust case}, the medium size {\tt nobel-eu} network was solved using the restricted-path version of STPH, for efficiency reasons. With the aim of evaluating the efficiency of STPH-RP  in Figure \ref{fig:STPH_vs_STPH-RP} the consumption obtained with both STPH and STPH-RP (with $\Omega = 10$) considering {\tt nobel-eu} network and \textit{dedicated classic} protection are reported. The consumption difference between the two solution methods is generally slightly lower than 10\%. Therefore, taking into account all the paths allows to gain a substantial amount of saving when handling small and medium size instances. However, the use of a restricted set of paths turned out to be a reasonable strategy to reduce the computational effort when dealing with larger instances without excessively degrading the achieved energy saving.

\begin{figure*}[t]\centering
  \includegraphics[width=18cm]{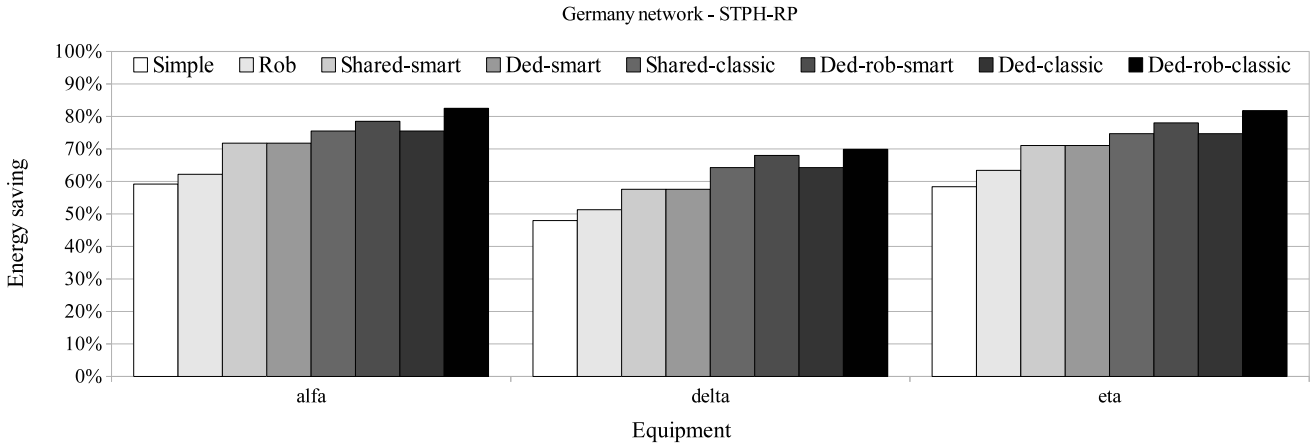}
  \caption{Energy savings achieved by STPH-RP with $\Omega\,=\,5$ when implementing the different protection schemes on {\tt germany} instances.}
\label{fig:germany_general}
\end{figure*}

Finally, to further confirm the good performance of STPH-RP, the average network consumption values computed on the {\tt germany} network with all the different protection schemes are reported in Figure~\ref{fig:germany_general}. We solved the instances with STPH-RP with $\Omega = 5$. Also in this case the heuristic method provides significant savings, obtaining final network consumption from 60\% up to 80\% of the original value.

\begin{table*}[!ht]\footnotesize
\tabcolsep 6pt
\caption{CPLEX time limits for the single time period to solve {\tt nobel-germany}, {\tt nobel-eu} and {\tt germany} instances with different types of protection.}
\begin{center}
\begin{tabular}{rcccccccccc}
& \multicolumn{2}{c}{\textit{simple}} & \multicolumn{2}{c}{\textit{robust}} & \multicolumn{2}{c}{\textit{dedicated}} & \multicolumn{2}{c}{\textit{shared}} & \multicolumn{2}{c}{\textit{robust-dedicated}} \\
\hline
Net & $TL_{STPH}$ & $TL_{STPH-RP}$ & $TL_{STPH}$ & $TL_{STPH-RP}$ & $TL_{STPH}$ & $TL_{STPH-RP}$ & $TL_{STPH}$ & $TL_{STPH-RP}$ & $TL_{STPH}$ & $TL_{STPH-RP}$ \\
\hline

nobel-ger	& 60s   & /    & 90s  & /    &  90s  & /    & 360s & /    & 120s  & /     \\
nobel-eu	& 300s  & /    & 300s & /    &  300s & 300s & /    & /    & 1200s & /     \\
germany  	& /     & 600s & /    & 600s &  /    & 600s & /    & /    & /     & 1200s \\
                                                                       
\hline 
\end{tabular}
\end{center}
\label{tab:time_limit}
\end{table*}

\section{Concluding Remarks}\label{sec:conclusions}
In this paper we have exhaustively investigated the issues concerning energy saving and network resilience to both failures and traffic variations. We have proposed a comprehensive set of modelling tools to efficiently perform multi-period off-line energy-aware network management without compromising the normal network operation. We have presented and discussed both exact and heuristic methods able to put to sleep network line cards and chassis while reserving backup resources to efficiently cope with single link failures and leaving enough spare capacity to absorb the unexpected peak of traffic. Extensive experimentations have shown that even when full protection is guaranteed (dedicated protection with robustness to traffic variations) it is possible to save up to 30\% of the daily network consumption.

\parpic{\includegraphics[width=.75in,height=1in]{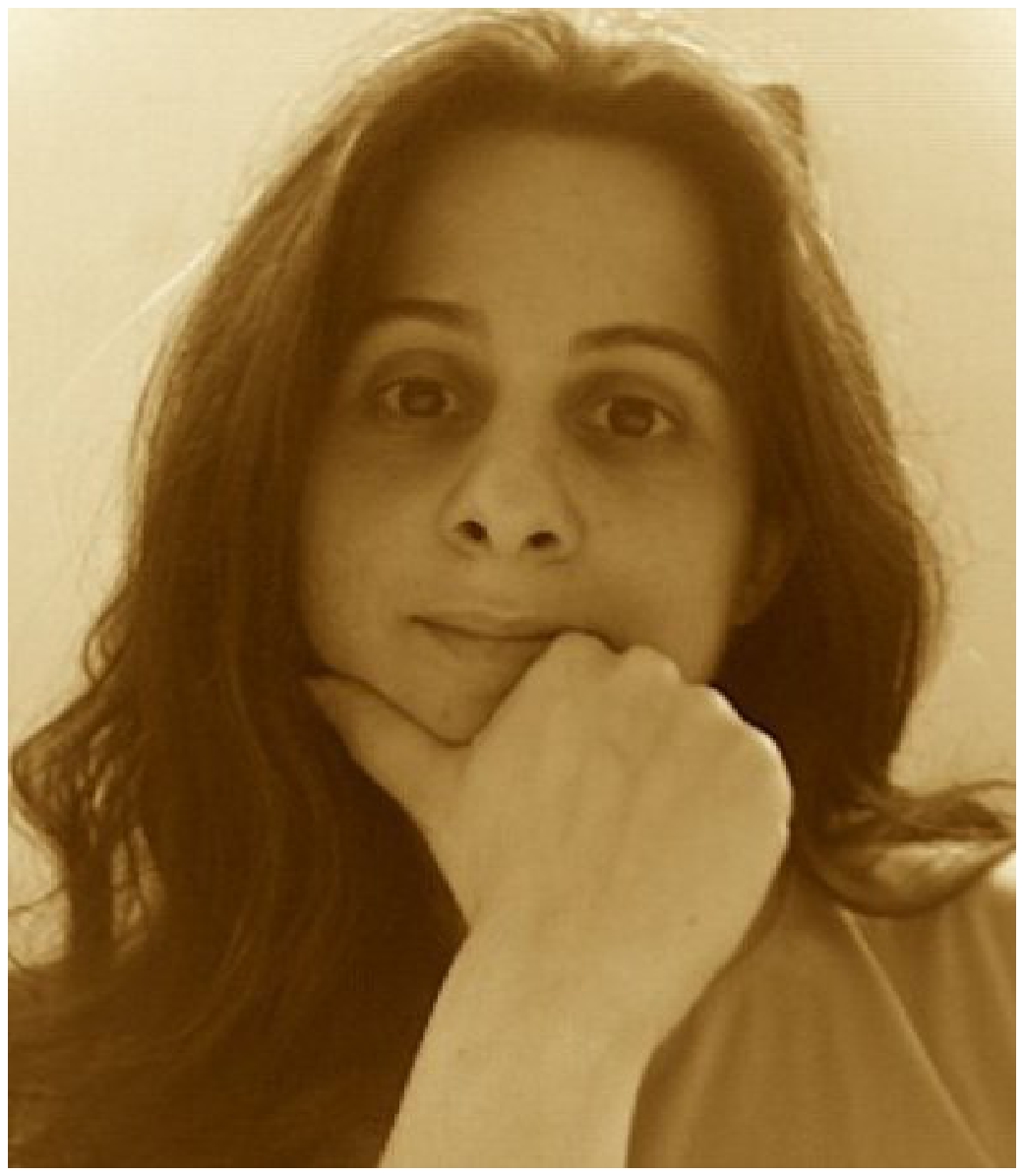}}{\bf Bernardetta Addis} is a Post-doc at the Computer Science Department (Dipartimento di Informatica) of the University of Turin (Universit\`a degli Studi di Torino), Her expertise is on optimization with particular reference to non linear global optimization and recently to heuristics and exact methods for integer optimization. She received the M.S. and Ph.D. degrees in computer science engineering from the University of Florence in 2001 and 2005, respectively.

\parpic{\includegraphics[width=1in,clip,keepaspectratio]{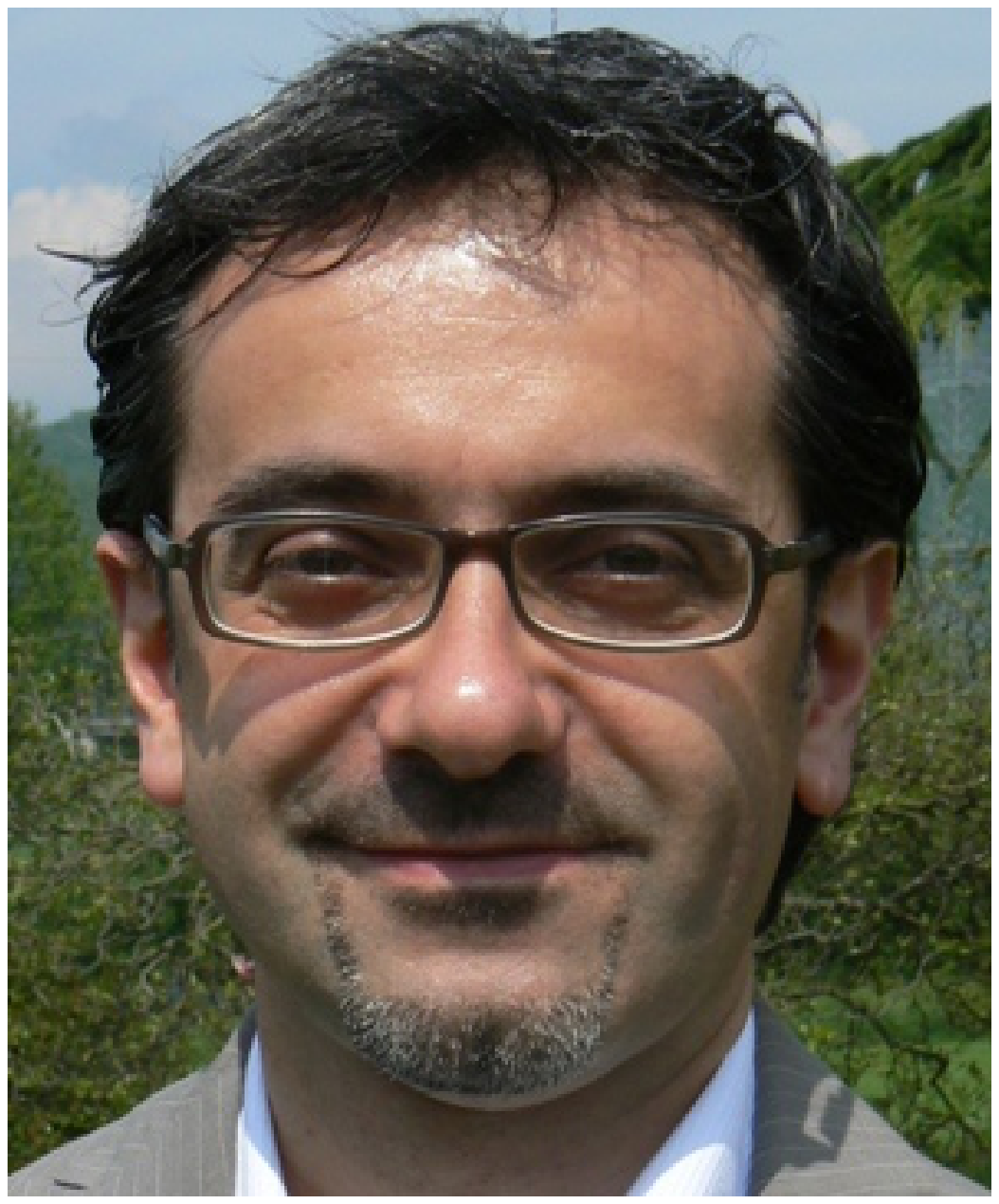}}{Antonio Capone}
\noindent {\bf Antonio Capone} is full professor at the Information and Communication Technology Department of the Politecnico di Milano (technical university), where he is the director of the Advanced Network Technologies Laboratory (ANTLab). His expertise is on networking and in this area he has published more than 200 peer-reviewed papers in international journals and conference proceedings.
He received the M.S. and Ph.D. degrees in electrical engineering from the Politecnico di Milano in 1994 and 1998, respectively. In 2000 he was visiting professor at UCLA, Computer Science department. He currently serves as editor of ACM/IEEE Trans. on Networking, Wireless Communications and Mobile Computing (Wiley), Computer Networks (Elsevier), and Computer Communications (Elsevier). He was guest editor of a few journal special issues  and served in the technical program committee of major international conferences. He is a Senior Member of the IEEE.

\parpic{\includegraphics[width=.75in,height=1in,clip,keepaspectratio]{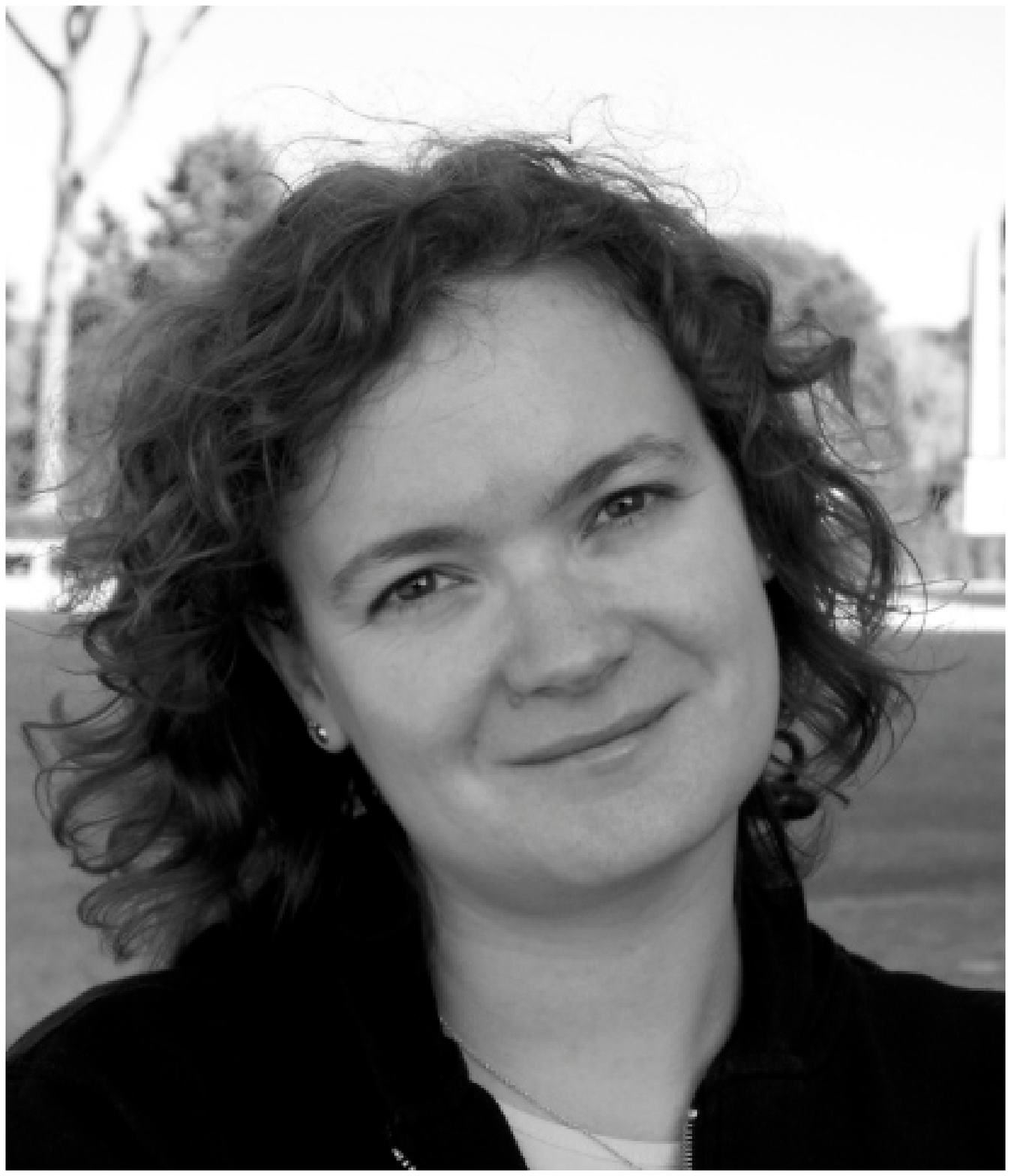}}{\bf Giuliana Carello} was born in 1972. She received her Laurea degree in 1999 from Politecnico di Torino. She earned her Ph.D. in 2004 in the same university, discussing the thesis ''Hub Location Problems in Telecommuncation Networks''. She visited Universit\'e  Libre de Bruxelles for six months in 2003. She joined the Operation Research Group of DEIB (Dipartimento di Elettronica, Informazione e Bioingegneria) of Politecnico di Milano as assistant professor in 2005. Her research work interests are exact and heuristic optimization algorithms, applied to integer and binary variable problems. Her research is mainly devoted to real life application: she focused on telecommunication network design problems. Besides the hub location problems, she worked on two layer wired network design problems and design and resource allocation problems in wireless networks. She published in international journals.

\parpic{\includegraphics[width=1in,clip,keepaspectratio]{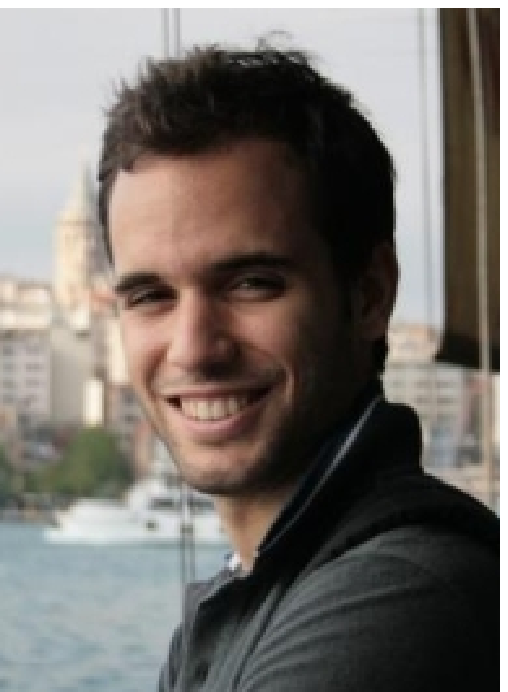}}\noindent {\bf Luca Giovanni Gianoli} was born in Milano in 1986. He received the Bachelor Degree in Telecommunications Engineering from Politecnico di Milano (Italy) in 2008, and the Master Degree in Telecommunications Engineering from Politecnico di Milano in 2010. Since January 2011 he's enrolled in a double Ph.D. program in Information Technology at Dipartimento di Elettronica, Informazione e Bioingegneria (DEIB) of Politecnico di Milano with Prof. Antonio Capone and D\'epartement de g\'enie \'electrique of \'Ecole Polytechnique de Montreal with Prof. Brunilde Sans\'o. His research interests include topics related to energy-aware network design and traffic engineering (green networking).

\parpic{\includegraphics[width=.75in,height=1in,clip,keepaspectratio]{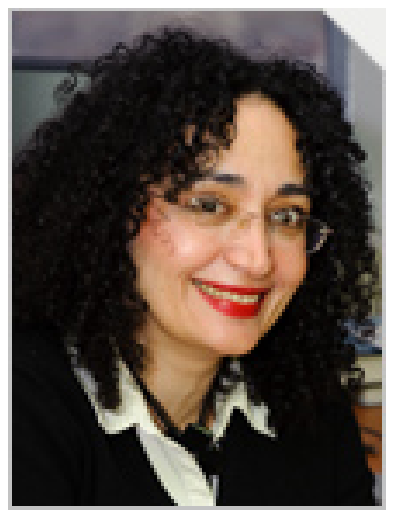}}]{\bf Brunilde Sans\'o} is a full professor of electrical engineering at \'Ecole Polytechnique de Montreal and director of the LORLAB. Her interests are in performance, reliability, design, and optimization of wireless and wireline networks. She is a recipient of several awards, Associate Editor of Telecommunication Systems, and editor of two books on planning and performance.

\end{document}